\documentclass[fleqn,usenatbib]{mnras}

\usepackage{newtxtext,newtxmath}

\usepackage[T1]{fontenc}

\DeclareRobustCommand{\VAN}[3]{#2}
\let\VANthebibliography\thebibliography
\def\thebibliography{\DeclareRobustCommand{\VAN}[3]{##3}\VANthebibliography}

\usepackage{graphicx}	
\usepackage{amsmath}	
\usepackage{color}
\usepackage{times}
\usepackage{hyperref}                  
\hypersetup{colorlinks=true, urlcolor=blue, linkcolor=blue,  citecolor=blue}
\usepackage{bm}
\usepackage{multirow}

\title[Ejecta velocity of SNe~Ia]{Calcium versus silicon ejecta velocities and decline rates in supernovae~Ia: The role of high-velocity features}

\author[A. A. Hakobyan et al.]{A.~A.~Hakobyan,$^{1}$\thanks{E-mail: \href{mailto:artur.hakobyan@yerphi.am}{artur.hakobyan@yerphi.am}}
M.~H.~Gevorgyan,$^{1}$
A.~G.~Karapetyan,$^{1}$
G.~A.~Mamon,$^{2}$
D.~Kunth,$^{2}$
\newauthor
V.~Adibekyan$^{3,4}$
and L.~V.~Barkhudaryan$^{1}$
\\
$^{1}$Center for Cosmology and Astrophysics, Alikhanian National Science Laboratory, 2 Alikhanian Brothers Str., 0036 Yerevan, Armenia\\
$^{2}$Institut d'Astrophysique de Paris (UMR 7095: CNRS and Sorbonne Universit\'{e}), 98 bis bd Arago, 75014 Paris, France\\
$^{3}$Instituto de Astrof\'isica e Ci\^encias do Espa\c{c}o, Universidade do Porto, CAUP, Rua das Estrelas, 4150-762 Porto, Portugal\\
$^{4}$Departamento de F\'{\i}sica e Astronomia, Faculdade de Ci\^encias, Universidade do Porto, Rua do Campo  Alegre, 4169-007 Porto, Portugal
}

\date{Accepted 2025 December 24. Received 2025 December 22; in original form 2025 August 19}

\pubyear{\the\year{}}

\begin{document}
\label{firstpage}
\pagerange{\pageref{firstpage}--\pageref{lastpage}}
\maketitle

\begin{abstract}
  Photospheric and high-velocity features (PVFs and HVFs) of \mbox{Si\,{\sc ii}}~$\lambda$6355 and \mbox{Ca\,{\sc ii}}~IR3
  lines in supernova Ia (SN Ia) spectra provide insights into ejecta structure, energetics, and circumstellar interaction,
  yet their interplay remains poorly understood.
  We analyse a representative sample of 145 nearby SNe~Ia observed within $\pm5$ days of $B$-band maximum light,
  including normal, 91T-, and 91bg-like events with measured light-curve decline rates $(\Delta m_{15})$ and
  \mbox{Si\,{\sc ii}} and \mbox{Ca\,{\sc ii}} line properties from the literature.
  We model PVF and HVF velocity distributions using Gaussian Mixture Models,
  compare \mbox{Si\,{\sc ii}} and \mbox{Ca\,{\sc ii}} PVF velocity distributions, assess \mbox{Ca\,{\sc ii}} HVF properties,
  and test correlations between \mbox{Si\,{\sc ii}} PVF velocities
  and $\Delta m_{15}$, with emphasis on HVF effects.
  For the first time, we show that the \mbox{Ca\,{\sc ii}} PVF velocity distribution,
  measured for the same events at the same phases as \mbox{Si\,{\sc ii}}, is predominantly unimodal,
  in contrast to the well-known bimodal \mbox{Si\,{\sc ii}} PVF distribution that supports
  the high-velocity/normal-velocity division.
  This contrast likely reflects a subclass-dependent formation
  depth of the \mbox{Ca\,{\sc ii}} line, as supported by a
  positive correlation $(>3.3\sigma)$ between $\Delta m_{15}$ and the velocity
  offset between \mbox{Ca\,{\sc ii}} and \mbox{Si\,{\sc ii}} PVFs,
  particularly in faster-declining SNe~Ia.
  Importantly, HVFs do not significantly bias PVF velocity distributions.
  A significant negative correlation $(>3.3\sigma)$ between \mbox{Si\,{\sc ii}} PVF velocity and $\Delta m_{15}$ is found only for HVF-weak SNe~Ia,
  consistent with more energetic explosions yielding faster ejecta,
  while this trend vanishes in HVF-strong events, likely due to circumstellar interaction.
  These results underscore the critical role of HVFs and SN~Ia subclass in
  interpreting ejecta kinematics in both models and observations.
\end{abstract}

\begin{keywords}
methods: data analysis -- supernovae: individual: Type Ia
\end{keywords}



\section{Introduction}
\label{intro}
\defcitealias{2015MNRAS.451.1973S}{S15}

Type Ia supernovae (SNe~Ia) are widely recognized as the thermonuclear explosions of
carbon-oxygen (CO) white dwarfs (WDs) in binary systems
\citep[e.g.][]{2000ARA&A..38..191H,2018PhR...736....1L}.
Owing to their high luminosities and relative uniformity, they serve as powerful cosmological distance indicators
\citep[e.g.][]{1998AJ....116.1009R,1999ApJ...517..565P}.
This utility stems from the empirical correlation between their peak luminosities and post-maximum light curve (LC) decline rates,
quantified by the $\Delta m_{15}$ parameter, the decrease in $B$-band magnitude during the first 15 days after maximum light
\citep{1993ApJ...413L.105P}.
Slower-declining SNe~Ia (with smaller $\Delta m_{15}$ values) tend to be more luminous,
while faster-declining events (higher $\Delta m_{15}$) are typically fainter.
Despite their success as standardizable candles, SNe~Ia exhibit considerable diversity in their spectral and kinematic properties,
reflecting a range of explosion energies, chemical stratification, and potentially distinct progenitor environments
\citep[see][for a recent review]{2025A&ARv..33....1R}.

Several explosion scenarios have been proposed to account for the diversity among SNe~Ia.
The canonical single-degenerate model involves an CO WD accreting material from a
non-degenerate companion, approaching the Chandrasekhar mass before ignition
\citep[][]{1973ApJ...186.1007W}.
Delayed detonation model, which introduces a deflagration-to-detonation transition,
enables a range of outcomes depending on ignition geometry and transition density
\citep[e.g.][]{1991A&A...245..114K,2007Sci...315..825M}.
Alternatively, in the double-degenerate model, merging of two WDs triggers the explosion
\citep[e.g.][]{1984ApJS...54..335I,2013ApJ...770L...8P}.
Sub-Chandrasekhar mass explosions, including double-detonation scenarios,
involve a He layer of varying thickness on a sub-$M_{\rm Ch}$ WD igniting a secondary detonation in the core
\citep[e.g.][]{2007ApJ...662L..95B,2021ApJ...922...68S}.
These models offer a natural framework for producing stratified ejecta,
in which radioactive $^{56}$Ni resides in the inner regions,
surrounded by intermediate mass element (IMEs) like silicon, sulfur, and calcium.

A defining observational feature of SNe~Ia is the presence of strong absorption lines in the spectra
from IMEs, notably \mbox{Si\,{\sc ii}}~$\lambda$6355 around 6150\text{\AA} and
\mbox{Ca\,{\sc ii}} near-infrared triplet (\mbox{Ca\,{\sc ii}}~IR3) around 8100\text{\AA}.
The velocities of these photospheric-velocity features (PVFs),
estimated from the blueshift of absorption minima,
trace the kinematics of expanding ejecta.

Much effort has been devoted to studying the \mbox{Si\,{\sc ii}} photospheric velocity distribution.
Notably, \citet{2013Sci...340..170W} introduced a bimodal classification into
normal-velocity (NV; $V_{\rm Si} < 12000$ km~s$^{-1}$) and high-velocity
(HV; $V_{\rm Si} \gtrsim 12000$ km~s$^{-1}$) events for spectroscopically normal SNe~Ia at maximum light
\citep[see also][]{2009ApJ...699L.139W,2020ApJ...895L...5P,2020MNRAS.499.5325Z}.
The physical nature of the two velocity groups is under active debate.
The bimodal velocity distribution has been interpreted as evidence for distinct progenitor or explosion channels
\citep[e.g.][]{2013Sci...340..170W,2014MNRAS.437..338C,2015MNRAS.446..354P,2019ApJ...873...84P,2020ApJ...895L...5P,2021ApJ...923..267D},
or as a manifestation of viewing-angle effects in asymmetric explosions
\citep[e.g][]{2010Natur.466...82M,2018MNRAS.477.3567M,2020MNRAS.499.5325Z}.

The \mbox{Ca\,{\sc ii}}~IR3 line properties and their correlations with various
photometric and spectroscopic observables have been examined in previous studies
\citep[e.g.][]{2014MNRAS.437..338C,2014MNRAS.444.3258M,2015MNRAS.446..354P}.
However, it remains unclear whether the bimodal velocity distribution observed in
the \mbox{Si\,{\sc ii}}~$\lambda$6355 line also extends to the \mbox{Ca\,{\sc ii}} photospheric component.
This question is best addressed by analysing both features at consistent maximum-light phases
(i.e. $t({\rm Si}) = t({\rm Ca})$).
Calcium forms at lower excitation energies (1.7~eV versus 8.12~eV for silicon),
making it sensitive to ejecta temperature structure and ionisation horizons
\citep[e.g.][]{2006MNRAS.370..299H}.
These differences can shift the effective line-formation region either deeper or outer within the ejecta
\citep[e.g.][]{2006MNRAS.370..299H,2015ApJS..220...20Z,2024MNRAS.535.3470Z},
potentially leading to velocity distributions that differ from those of the silicon line.
Whether the bimodal behaviour also extends to the \mbox{Ca\,{\sc ii}} velocities
therefore remains an open question.

Spectroscopic diversity extends beyond velocity differences.
In addition to normal SNe~Ia \citep[e.g.][]{1993AJ....106.2383B},
several spectral subclasses have been established,
including the overluminous 91T-like events \citep[e.g.][]{1992ApJ...384L..15F,1992AJ....103.1632P}
and the subluminous 91bg-like subclass \citep[e.g.][]{1992AJ....104.1543F,1993AJ....105..301L}.
91T-like SNe~Ia are characterized by weak IME lines at early phases and slow
LC decline rates ($\Delta m_{15} \lesssim 1$),
while 91bg-like events show prominent \mbox{Ti\,{\sc ii}} absorption and
fast-declining LCs ($\Delta m_{15} \gtrsim 1.6$).
Normal SNe~Ia populate the intermediate range in both photometric and spectroscopic
properties.\footnote{For more details on the properties of these events,
as well as other transitional SNe~Ia, the reader is referred to \citet{2017hsn..book..317T}.}
Understanding how these subclasses differ in their explosion energetics and chemical stratification
is essential for identifying the progenitor systems and explosion mechanisms of SNe~Ia
\citep[e.g.][]{2000ARA&A..38..191H,2018PhR...736....1L,2025A&ARv..33....1R}.

High-velocity features (HVFs), commonly observed in the early spectra of SNe~Ia,
particularly in \mbox{Ca\,{\sc ii}}~IR3 and \mbox{Ca\,{\sc ii}}~H\&K
(around 3700\text{\AA}) lines, add an additional layer of complexity.
These HVFs manifest as detached absorption components at velocities exceeding 18000 km~s$^{-1}$,
often appearing in the pre-maximum phase and fading rapidly afterwards
(e.g. \citealt{2014MNRAS.437..338C,2014MNRAS.444.3258M,2015ApJS..220...20Z};
\citealt{2015MNRAS.451.1973S}, hereafter S15).
While the origin of HVFs is still debated, proposed mechanisms include interaction with circumstellar material (CSM),
density or abundance enhancements in the outer ejecta, or shell-like structures formed during the explosion
\citep[e.g.][]{2005MNRAS.357..200M,2008ApJ...677..448T,2019MNRAS.484.4785M,2019ApJ...886...58M,2025arXiv251213791H}.
Notably, the strength of HVFs varies significantly across SN~Ia subclasses and
may be linked to progenitor age and environment
(e.g. \citealt{2014MNRAS.437..338C,2014MNRAS.444.3258M,2015MNRAS.446..354P,2015ApJS..220...20Z};
\citetalias{2015MNRAS.451.1973S};
\citealt{2019ApJ...886...58M}).

Changes in SN~Ia LC decline rate, parameterised by $\Delta m_{15}$, offer an indirect probe of
the explosion energy and thermal conditions \citep[e.g.][]{2005ApJ...623.1011B,2006MNRAS.370..299H}.
More energetic explosions, typically associated with low $\Delta m_{15}$ (i.e. slow-declining events),
are expected to yield faster-expanding ejecta.
However, contrary to theoretical expectations, recent studies found only weak
trends between photospheric velocities and $\Delta m_{15}$
\citep[e.g.][]{2014MNRAS.437..338C,2014MNRAS.444.3258M,2021ApJ...923..267D,2021MNRAS.503.4667Z,2024MNRAS.532.1887P}.
Interestingly, \citet{2014MNRAS.437..338C} showed that this trend might be environment-dependent.
This points to additional factors, such as HVFs, potentially modulating the ejecta velocity structure.

In this study, we aim to advance the understanding of the coupling between SN~Ia ejecta kinematics,
photometric diversity, and the potential role of progenitor and environmental influences.
To this end, we perform a unified and phase-consistent ($t({\rm Si}) = t({\rm Ca})$) analysis of
the PVFs of \mbox{Si\,{\sc ii}}~$\lambda$6355 and \mbox{Ca\,{\sc ii}}~IR3 lines,
as well as the HVFs of the \mbox{Ca\,{\sc ii}}~IR3,
using a representative sample of 145 nearby SNe~Ia.
All velocity measurements are obtained within a phase interval of $\pm5$ days centered on $B$-band maximum light.
The sample spans a broad range of LC decline rates and includes spectroscopically classified normal,
91T-, and 91bg-like events.

Our paper is structured as follows.
Section~\ref{samplered} presents the sample selection.
In Section~\ref{RESsDIS1}, we examine the velocity distributions of PVFs from
\mbox{Si\,{\sc ii}}~$\lambda$6355 and \mbox{Ca\,{\sc ii}}~IR3,
highlighting key differences in their profiles and subclass dependencies.
In Section~\ref{RESsDIS2}, we explore the statistical behavior of \mbox{Ca\,{\sc ii}} HVFs,
their strengths, and the impact of HVFs on the distributions of PVF velocities.
In Section~\ref{RESsDIS3}, we investigate the relationship between \mbox{Si\,{\sc ii}} PVF velocities
and LC decline rates, placing emphasis on the role of HVFs in modulating observed trends.
Finally, Section~\ref{CONCs} summarizes our conclusions.

\section{Sample selection}
\label{samplered}

All the data used in this study, concerning the Si and Ca absorption
features in spectra of SNe~Ia, are compiled from \citetalias{2015MNRAS.451.1973S}.
A total of 445 spectra, obtained earlier than 5~days after maximum brightness for
210 SNe~Ia with $z < 0.1$ (median is 0.02), are included in the dataset of
\citetalias{2015MNRAS.451.1973S}.
Based on photometric observations and the carefully determined maximum brightness of these SNe~Ia,
each spectrum has a well-defined phase (see details in \citetalias{2015MNRAS.451.1973S}).
Adopting the optically thin approximation,
a single Gaussian and three superposed Gaussian functions were employed by
\citetalias{2015MNRAS.451.1973S} to find the best fit for the \mbox{Si\,{\sc ii}}~$\lambda$6355 and
\mbox{Ca\,{\sc ii}}~IR3 absorption features, respectively.
Moreover, for each feature, two separate fits were applied: PVF and HVF fits.
Note that, HVFs undergo a rapid decline following the SN explosion.
In the case of \mbox{Si\,{\sc ii}}~$\lambda$6355,
these features become significantly weakened approximately one week prior to
maximum brightness.\footnote{\citet{2025A&A...695A.264H} reported that \mbox{Si\,{\sc ii}} HVFs
may still be detectable at moderate level in up to one-third of spectra obtained
within one week prior to maximum light.}
However, in the \mbox{Ca\,{\sc ii}}~IR3, HVFs remain detectable at near-maximum light and
can persist for several days beyond this phase (e.g.
\citealt{2013ApJ...777...40M,2013ApJ...770...29C,2014MNRAS.437..338C};
\citetalias{2015MNRAS.451.1973S}).
The velocity\footnote{\citetalias{2015MNRAS.451.1973S}
reported uncertainties of $\sim200-400$~${\rm km~s}^{-1}$ in their velocity calculations,
based on the accuracy of the fits.}
and pseudo-equivalent width (pEW) of each feature were subsequently calculated using the fit parameters,
the relativistic Doppler formula, and a predefined spectral
pseudo-continuum.\footnote{Note that \citetalias{2015MNRAS.451.1973S} also measured
the \mbox{Ca\,{\sc ii}}~H\&K absorption feature,
however, we did not include this data in the current study due to the potential risk of blending
with \mbox{Si\,{\sc ii}}~$\lambda$3858 absorption \citep[e.g.][]{2013MNRAS.435..273F},
which would complicate further interpretation.}

Our study focuses on near-maximum light phases,
during which both the PVF and HVF components of the \mbox{Si\,{\sc ii}} and \mbox{Ca\,{\sc ii}}
lines have been measured simultaneously ($t({\rm Si}) = t({\rm Ca})$).
For further comparative analysis,
we followed the approach of \citet{2014MNRAS.444.3258M} and limited
the data to a narrow phase interval where the velocity evolution of the PVFs is relatively small
\citep[see also][]{2025A&A...694A...9B}.
Accordingly, we selected SNe~Ia from \citetalias{2015MNRAS.451.1973S}
that meet the specified criteria $t({\rm Si}) = t({\rm Ca})$
and are within a ten-day phase interval centered at $t=0$ day.
This formed our SN sample.

SNe Ia in the dataset of \citetalias{2015MNRAS.451.1973S} are spectroscopically subclassified
using the well-established SuperNova IDentification algorithm \citep[SNID;][]{2007ApJ...666.1024B}.
Based on this classification data, the SNe~Ia that meet the criteria for our sample
comprise 142 SNe~Ia with reliable subclassifications,
while three events were simply classified as Type Ia SNe rather than
being further subclassified.
For the $B$-band LC decline rates $(\Delta m_{15})$ of the 145 SNe~Ia,
\citetalias{2015MNRAS.451.1973S} were able to collect data
for 89 events from different published sources.

For the SN sample in this study, we aimed to maximize the dataset size by including all available
$\Delta m_{15}$ values and spectroscopic subclassifications.
Therefore, based on our earlier experience \citep[e.g.][]{2019MNRAS.490..718B,2020MNRAS.499.1424H},
we conducted a thorough literature search to compile additional (and more precise) information on
the subclassification and the $B$-band LC data for the SNe~Ia under study.
First, we cross-matched these SNe~Ia with our database \citep{2020MNRAS.499.1424H},
which includes, in particular, detailed information on the spectroscopic subclasses of
407 nearby SNe~Ia and their $B$-band $\Delta m_{15}$ values.
For SNe~Ia in common, we adopted LC decline rates and subclasses from \citet{2020MNRAS.499.1424H}.
In addition, for the SNe~Ia with lacking $\Delta m_{15}$ values,
we collected the data from a variety of publications in which
different LC fitters were used on $B$-band photometry with broad temporal coverage
\citep[e.g.][]{2012AJ....143..126B,2018ApJ...867...56B,2018PASP..130k4504P}.\footnote{All adopted
$\Delta m_{15}$ values are obtained either from polynomial fits or from
template-based LC fitting methods applied to $B$-band photometry.
No stretch-based LC fitters are used to derive $\Delta m_{15}$,
ensuring that the decline-rate measurements are robust for all SN~Ia subclasses, including 91bg-like events.}
All $\Delta m_{15}$ values include their published uncertainties,
and these uncertainties are incorporated into our Monte Carlo (MC) approach when
evaluating the robustness of the derived correlations in this study.

After these operations, our final sample comprises 113 spectroscopically normal SNe~Ia,
15 of the 91T-like subclass, and 17 of the 91bg-like subclass,
all with $z \lesssim 0.08$ (median is 0.02).
Note that overluminous 91T-like subclass includes also 99aa-like SNe~Ia
with close spectral and photometric characteristics.
Among these SNe~Ia, 111 have velocity measurements at a single phase within the ten-day interval,
18 have measurements at two distinct phases, and 16 have data points obtained at three or more
different phases within the same interval.
In total, we were able to obtain LC decline rates for 121
events out of 145 SNe~Ia, which is an additional
32 SNe with available $\Delta m_{15}$ values
compared to the initial \citetalias{2015MNRAS.451.1973S} data.
Table~\ref{SNIaNum} presents the distribution of SN~Ia subclasses in our sample.

\begin{table}
  \caption{Distribution of SN~Ia subclasses in our sample $(-5 \leq t \leq +5)$.}
  \label{SNIaNum}
  \centering
  \tabcolsep 15.4pt
      \begin{tabular}{lccc}
    \hline
  \multicolumn{1}{l}{SN} && \multicolumn{1}{c}{$N_{\rm SN}$} & \multicolumn{1}{c}{$N_{\rm SN}$ with available $\Delta m_{15}$} \\
  \hline
   Normal && 113 & 97 \\
   91T && 15 & 12 \\
   91bg && 17 & 12 \\
   All && 145 & 121 \\
  \hline
  \end{tabular}
\end{table}

We note that, adopting the optically thick approximation and a different line-fitting approach,
\citet{2015ApJS..220...20Z} were able to measure the PVFs and HVFs of
the \mbox{Si\,{\sc ii}}~$\lambda$6355 and \mbox{Ca\,{\sc ii}}~IR3 lines for
204 SNe~Ia with $z < 0.08$ at various phases.
However, we preferred not to combine the line measurements from \citetalias{2015MNRAS.451.1973S} and \citet{2015ApJS..220...20Z},
since these two studies are based on different physical assumptions and fitting constraints,
and their combination could introduce systematic inconsistencies.
Moreover, in contrast to \citetalias{2015MNRAS.451.1973S},
\citet{2015ApJS..220...20Z} did not attempt to measure
the HVFs for the \mbox{Si\,{\sc ii}} line at near-maximum light.
Although the use of more recent SN~Ia samples is highly desirable,
spectral measurements meeting the above mentioned criteria remain scarce in the literature,
and the current selection in Table~\ref{SNIaNum} represents the best available choice.

It is important to assess the representativeness of the SN~Ia subclasses in our sample relative to those in a nearly complete,
volume-limited sample, such as the Lick Observatory Supernova Search \citep[LOSS;][]{2011MNRAS.412.1441L} with redshifts $z \lesssim 0.02$.
In our sample, the subclass fractions of normal, 91T-, and 91bg-like SNe~Ia are
$78^{+3}_{-4}$, $10^{+3}_{-3}$, and $12^{+3}_{-3}$ per cent, respectively.
For comparison, the corresponding fractions in LOSS are
$74^{+6}_{-6}$, $10^{+5}_{-4}$, and $16^{+6}_{-5}$ per cent.
The values are also consistent with those of our previous SN~Ia sample with $z \lesssim 0.04$ \citep{2020MNRAS.499.1424H},
where the subclass fractions are $77\pm2$, $11\pm2$, and $13\pm2$ per cent
for normal, 91T-, and 91bg-like events, respectively.
More recently, \citet{2025A&A...698A.305M} constructed a volume-limited, nearly complete sample of
SNe in the nearby Universe discovered primarily by modern wide-field untargeted surveys between 2016 and 2023,
finding subclass fractions of $78^{+5}_{-6}$, $6^{+5}_{-3}$, and $16^{+6}_{-5}$ per cent for normal, 91T-, and 91bg-like SNe~Ia, respectively.
Given the close agreement among all these datasets,
we conclude that our sample is representative of the general SN~Ia population,
and that any potential subclass-related selection biases are not statistically
significant.\footnote{We refrained from including SNe~Ia from higher-redshift surveys designed for cosmological use,
as these often excluded 91bg-like and fast-declining normal events.
Including such samples could substantially alter the subclass distribution and
compromise the representativeness of our SN~Ia sample.}

\section{Results and discussion}
\label{RESsDIS}

In this section, we conduct a detailed investigation of the PVFs traced by
the \mbox{Si\,{\sc ii}}~$\lambda$6355 and \mbox{Ca\,{\sc ii}}~IR3 absorption lines in SNe~Ia,
alongside their LC decline rates.
Particular emphasis is placed on characterizing the HVFs associated with the \mbox{Ca\,{\sc ii}}~IR3 line.

\subsection{Distributions of PVF velocities}
\label{RESsDIS1}

Following SN~Ia explosion, the outer layers of the expanding ejecta
are primarily composed of IMEs, such as Si and Ca.
Since Si is the most abundant IME at the photospheric phase
\citep[e.g.][]{2007Sci...315..825M},
the expansion velocity $(V_{\rm Si})$ of the bulk ejecta is typically measured from
the blueshifted absorption minimum of the strongest
\mbox{Si\,{\sc ii}} optical line around 6150\text{\AA}.
Similarly, another strong \mbox{Ca\,{\sc ii}} line in the near IR band
(around 8100\text{\AA}) can be utilized to estimate the ejecta's expansion
velocity $(V_{\rm Ca})$.
The velocities, which are proportional to the radii of absorbing layers,
drop over time as the SN ejecta expands, exposing deeper layers \citep[e.g.][]{2011ApJ...742...89F}.
This is a characteristic temporal tendency for the PVF velocities $(V^{\rm PVF})$
of the Si and Ca lines (e.g. \citealt{2012MNRAS.425.1819S}; \citetalias{2015MNRAS.451.1973S}).

To study the distribution of expansion velocities,
$V_{\rm Si}$ and $V_{\rm Ca}$ values are usually converted,
using various average best-fitting curves, to their values at $B$-band maximum light,
denoted as $V^0_{\rm Si}$ and $V^0_{\rm Ca}$, respectively
\citep[e.g.][]{2011ApJ...742...89F,2013Sci...340..170W,2020MNRAS.499.5325Z}.
However, detailed phase-resolved studies of individual SNe~Ia have revealed a wide range of velocity drop over time,
including exponential, linear, and other non-linear evolutions
(e.g. \citealt{2005ApJ...623.1011B,2011ApJ...742...89F,2012AJ....143..126B,2012MNRAS.425.1819S};
\citetalias{2015MNRAS.451.1973S}; \citealt{2025A&A...695A.264H}).
The converted velocities include various uncertainties due to the different intrinsic phase evolution.
Therefore, following \citet{2014MNRAS.444.3258M},
we do not correct the velocities to those of maximum brightness for our analysis
(or when linking the velocities with the SN LC decline rates).
Instead, we limit the data to a ten-day phase interval
centered at $t=0$ day.

Importantly, \citet{2013Sci...340..170W} demonstrated that the distribution of
\mbox{Si\,{\sc ii}}~$\lambda$6355 expansion velocities
of SNe~Ia near maximum light consists of two distinct groups:
NV $(< 12000$ km~s$^{-1})$ and HV $(\gtrsim 12000$ km~s$^{-1})$ events.
Furthermore, statistical analyses have shown that a bimodal Gaussian function
(Gaussian Mixture Model) provides the best maximum-likelihood (ML) fit to
the observed Si velocity distribution, significantly outperforming a unimodal fit
\citep[][]{2020MNRAS.499.5325Z}.
The observed velocity bimodality has been interpreted in terms of differing
progenitor natures \citep[e.g.][]{2013Sci...340..170W,2015MNRAS.446..354P},
explosion scenarios \citep[e.g.][]{2005ApJ...623.1011B,2006MNRAS.370..299H,2014MNRAS.437..338C,2019ApJ...873...84P},
or as a consequence of asymmetric explosions,
wherein the observed ejecta velocity variations depend on the viewing angle
\citep[e.g][]{2010Natur.466...82M,2018MNRAS.477.3567M,2020MNRAS.499.5325Z}.
Despite extensive work on \mbox{Si\,{\sc ii}} velocities,
the corresponding distribution of \mbox{Ca\,{\sc ii}}~IR3 velocities,
particularly those measured at the same phases as the \mbox{Si\,{\sc ii}}~$\lambda$6355 line,
has not yet been systematically examined.
Addressing this gap is essential for understanding the formation properties of one of
the most prominent near-infrared absorption features in SN~Ia spectra.

\begin{figure*}
\begin{center}$
\begin{array}{@{\hspace{0mm}}c@{\hspace{2mm}}c@{\hspace{0mm}}}
\includegraphics[width=0.42\hsize]{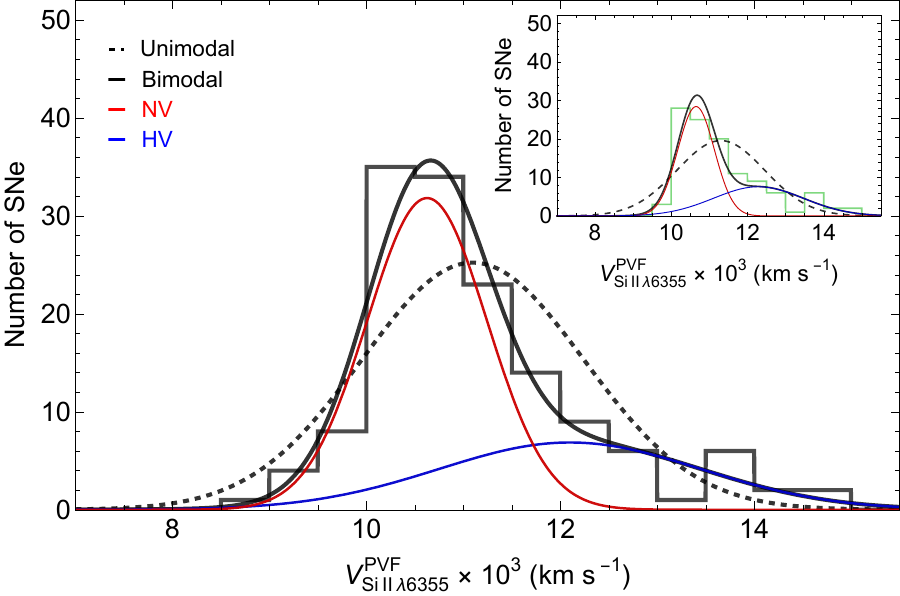} & \includegraphics[width=0.43\hsize]{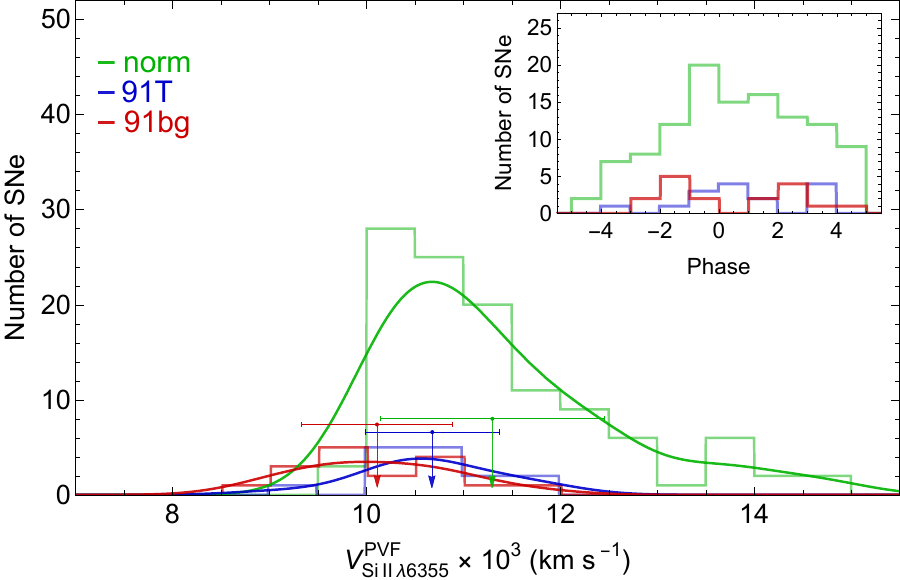}\\
\includegraphics[width=0.42\hsize]{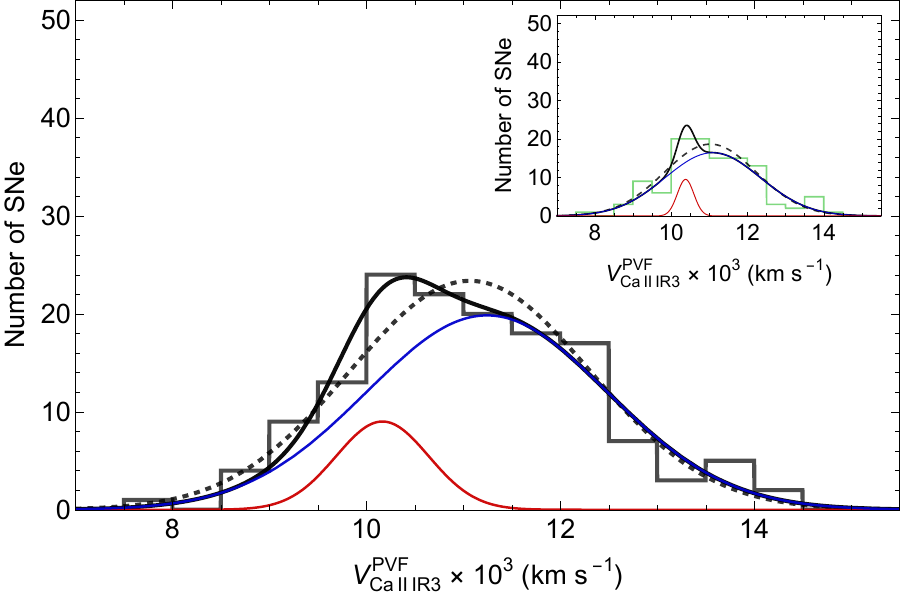} & \includegraphics[width=0.43\hsize]{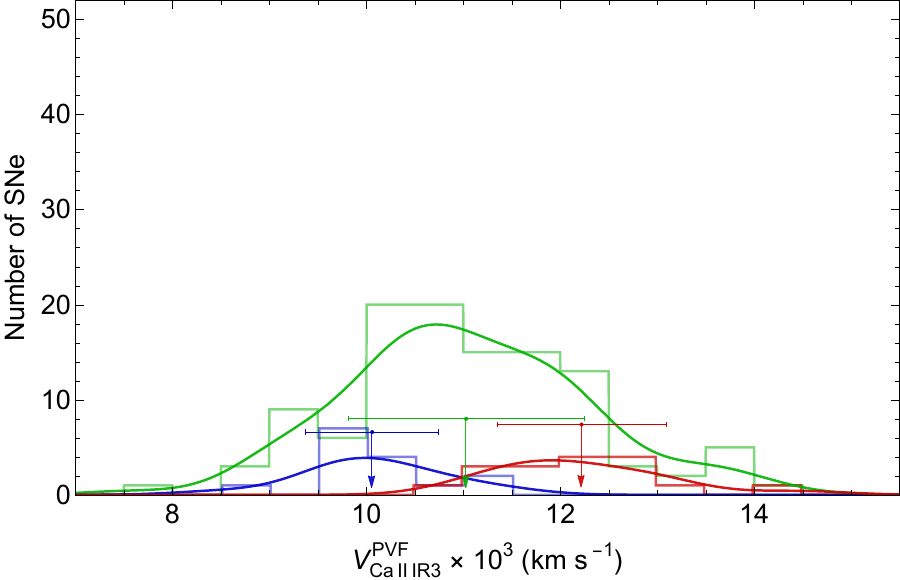}\\
\end{array}$
\end{center}
\caption{\emph{Left:} Distributions of PVF velocities for the \mbox{Si\,{\sc ii}}~$\lambda$6355 and \mbox{Ca\,{\sc ii}}~IR3 lines
         for all SN~Ia subclasses (black histograms).
         The bimodal and unimodal Gaussian fits to the distributions are presented by solid and dashed thick lines, respectively.
         The NV and HV components of the bimodal fit are indicated by red and blue thin lines, respectively.
         The insets display the same distributions and corresponding fits, but restricted to normal SNe~Ia (green histograms).
         \emph{Right:} Distributions of $V^{\rm PVF}_{\rm Si}$ and $V^{\rm PVF}_{\rm Ca}$,
         with overlaid smoothed histograms, separated by SN~Ia subclasses:
         normal (green), 91T- (blue), and 91bg-like (red) SNe~Ia.
         Arrows indicate the means of the distributions, with their standard deviations shown as horizontal error bars.
         The inset presents the histograms of phases at which the velocity measurements were obtained for each SN~Ia subclass.}
\label{SNVeldis}
\end{figure*}

In the left panels of Fig.~\ref{SNVeldis}, we present the distributions of
$V^{\rm PVF}_{\rm Si}$ and $V^{\rm PVF}_{\rm Ca}$, along with their bimodal and unimodal Gaussian fits,
for all SN~Ia subclasses in our sample.
All the fits, both bimodal and unimodal, are performed using a maximum-likelihood estimation (MLE) technique,
following the formalism outlined in Sections~3.1--3.2 of \citet{2020MNRAS.499.5325Z},
implemented in the \emph{Wolfram Mathematica} environment.
For each bimodal and unimodal Gaussian model, we maximise the total log-likelihood with respect to all free parameters
(means, standard deviations, and, for the bimodal case, the component mixing proportion),
allowing the algorithm to converge without additional constraints.
This likelihood-based approach avoids the binning sensitivity inherent in histogram fitting and
yields statistically robust and internally consistent parameter estimates for the fits.

The upper insets in the left panels of Fig.~\ref{SNVeldis} display the same distributions and fits,
but restricted to normal SNe~Ia.
The right panels of Fig.~\ref{SNVeldis} show the distributions of $V^{\rm PVF}_{\rm Si}$ and $V^{\rm PVF}_{\rm Ca}$,
with smoothed histograms overlaid for visualization, separated by SN~Ia subclassifications:
normal, 91T-, and 91bg-like SNe~Ia.
The inset of the upper right panel presents the histograms of phases at which
velocity measurements were obtained for each SN~Ia subclass.

In Table~\ref{VelPVFdisTest}, we present the best-fit MLE parameters for both
bimodal and unimodal Gaussian models applied to the velocity distributions.
To assess the statistical improvement of the bimodal fit over the unimodal fit,
we conducted one-sample Kolmogorov–Smirnov (KS) and Anderson–Darling (AD)
tests\footnote{Compared to the KS test, the AD test is more sensitive to differences
in the tails of distributions \cite[e.g.][]{Engmann+11}.} on the velocity datasets.
We define the null hypothesis as the assumption that the data follow the fitted bimodal (unimodal) Gaussian distribution,
whereas the alternative hypothesis asserts that the data do not originate from the specified Gaussian model.
In Table~\ref{VelPVFdisTest},
the $P$-value obtained from the one-sample (KS or AD) test represents the probability that the observed velocity distribution
is randomly drawn from the fitted parent distribution.
Following established statistical conventions, we adopt a significance threshold of $P = 0.05$ throughout this study
to determine whether the null hypothesis can be rejected.

\begin{table*}
  \caption{Best-fit parameters for both bimodal and unimodal Gaussian models applied to
           the \mbox{Si\,{\sc ii}} and \mbox{Ca\,{\sc ii}} PVF velocity $(\times 10^3)$ distributions of all SNe~Ia and normal events.
           The $P$-values from the KS and AD tests,
           which assess the goodness-of-fit between the observed velocity data and each model,
           along with the corresponding AICc and median $\Delta$AICc values used for model selection, are also listed.}
  \label{VelPVFdisTest}
  \centering
  \tabcolsep 3.75pt
    \begin{tabular}{lccccccccccccccccc}
    \hline
   & & \multicolumn{9}{c}{---------------------------------------- Bimodal ---------------------------------} & \multicolumn{5}{c}{----------------- Unimodal -----------------} && \\
   & & \multicolumn{3}{c}{------ NV Group ------} & \multicolumn{3}{c}{------ HV Group ------} &&&&&&&&&& \\
  \multicolumn{1}{l}{SN} & \multicolumn{1}{c}{$N$} & \multicolumn{1}{c}{$\mu_1$} & \multicolumn{1}{c}{$\sigma_1$} & \multicolumn{1}{c}{$p$} & \multicolumn{1}{c}{$\mu_2$} & \multicolumn{1}{c}{$\sigma_2$} & \multicolumn{1}{c}{$1-p$} & \multicolumn{1}{c}{$P_{\rm KS}^{\rm MC}$} & \multicolumn{1}{c}{$P_{\rm AD}^{\rm MC}$} & \multicolumn{1}{c}{AICc} & \multicolumn{1}{c}{$\mu$} & \multicolumn{1}{c}{$\sigma$} & \multicolumn{1}{c}{$P_{\rm KS}^{\rm MC}$} & \multicolumn{1}{c}{$P_{\rm AD}^{\rm MC}$} & \multicolumn{1}{c}{AICc} & \multicolumn{1}{c}{$\Delta$AICc} & \multicolumn{1}{c}{Med($\Delta$AICc)} \\
   && \multicolumn{1}{c}{(km/s)} & \multicolumn{1}{c}{(km/s)} & \multicolumn{1}{c}{(\%)} & \multicolumn{1}{c}{(km/s)} & \multicolumn{1}{c}{(km/s)} & \multicolumn{1}{c}{(\%)} &&&& \multicolumn{1}{c}{(km/s)} & \multicolumn{1}{c}{(km/s)} &&&& \\
   \hline
  & & \multicolumn{14}{c}{------------------------------------------------------- Si PVF velocities --------------------------------------------------------} && \\
   All & 145 & 10.6 & 0.6 & 68$\pm$4 & 12.1 & 1.3 & 32$\pm$4 & 0.978 & 0.968 & 426.8 & 11.1 & 1.1 & \textbf{0.028} & \textbf{0.013} & 455.2 & \textbf{28.4} & $\textbf{23.1}^{\textbf{+5.0}}_{\textbf{--4.6}}$ \\
   Normal & 113 & 10.6 & 0.5 & 60$\pm$5 & 12.3 & 1.2 & 40$\pm$5 & 0.877 & 0.885 & 325.5 & 11.3 & 1.1 & \textbf{0.031} & \textbf{0.013} & 356.9 & \textbf{31.4} & $\textbf{23.2}^{\textbf{+5.3}}_{\textbf{--5.0}}$ \\
   & & \multicolumn{14}{c}{------------------------------------------------------- Ca PVF velocities --------------------------------------------------------} && \\
   All & 145 & 10.2 & 0.5 & 15$\pm$4 & 11.2 & 1.2 & 85$\pm$4 & 0.988 & 0.927 & 482.5 & 11.1 & 1.2 & 0.892 & 0.927 & 477.4 & 5.1 & $2.5^{+2.1}_{-1.7}$ \\
   Normal & 113 & 10.4 & 0.2 & 9$\pm$3 & 11.1 & 1.2 & 91$\pm$3 & 0.997 & 0.977 & 482.2 & 11.0 & 1.2 & 0.853 & 0.874 & 477.8 & 4.5 & $2.6^{+2.0}_{-1.7}$ \\
  \hline
  \end{tabular}
  \parbox{\hsize}{\emph{Notes.} For the one-sample KS and AD tests, MC simulations with $10^4$ iterations are used to provide
  $P_{\rm KS}^{\rm MC}$ and $P_{\rm AD}^{\rm MC}$ probabilities.
  Bold-marked $P$-values $(\leq0.05)$ indicate statistically significant differences between the observed data and the fitted model.
  AICc differences greater than 6 are highlighted in bold,
  representing strong evidence in favor of the model with the lower AICc value.}
\end{table*}

From another perspective, to independently assess the statistical improvement of one fit over another,
we follow \citet{2020MNRAS.499.5325Z} and employ the corrected Akaike Information Criterion
\citep[AICc;][]{1974ITAC...19..716A,Sugiura1978,HurvichTsai89}
as a model selection method.
In particular, AICc value is derived from three key components:
model complexity ($k$, the number of free parameters used in model construction),
sample size $(N)$, and the ML estimate $(L_{\rm max})$,
which quantifies how well the model represents the observed data:
\begin{flalign*}
& {\rm AICc} = 2k - 2\log L_{\rm max} + \dfrac{2k(k+1)}{N-k-1} \, .&
\end{flalign*}
In this context, a unimodal Gaussian distribution
$\mathcal{N}(\mu, \sigma)$
is defined by two parameters:
the mean $(\mu \equiv \langle V \rangle)$ and standard deviation $(\sigma)$, resulting in a model complexity of 2.
In contrast, a bimodal Gaussian distribution
$p\mathcal{N}(\mu_1, \sigma_1) + (1-p)\mathcal{N}(\mu_2, \sigma_2)$
is characterized by five parameters:
the mixing proportion ($p$, weight of the first normal component),
the mean $(\mu_1 \equiv \langle V \rangle_{\rm NV})$ and standard deviation $(\sigma_1)$ of the first normal distribution,
and the mean $(\mu_2 \equiv \langle V \rangle_{\rm HV})$ and standard deviation $(\sigma_2)$ of the second normal distribution
(see Table~\ref{VelPVFdisTest}), resulting in a model complexity of 5.
While more complex models generally provide better fits to the data,
they also increase the risk of overfitting.
To mitigate this, AICc applies a penalty for additional parameters,
ensuring a balance between model complexity and goodness of fit.
The difference between AICc values determines model preference:
\begin{flalign*}
& \Delta \, {\rm AICc} = {\rm AICc}_{\rm higher} - {\rm AICc}_{\rm lower} \, .&
\end{flalign*}
An AICc difference of greater than 6 provides strong positive evidence
in favor of the model with the lower AICc value.
Conversely, if the difference is smaller, it suggests that the additional complexity of
the bimodal Gaussian model is not justified
(for further details, see \citealt{2020MNRAS.499.5325Z} and references therein).
In Table~\ref{VelPVFdisTest}, we present the AICc values for both unimodal and bimodal models
and their $\Delta$AICc difference.

For the PVF velocity distribution of the \mbox{Si\,{\sc ii}} line,
we find that the bimodal Gaussian model provides a significantly better fit to the data,
whereas the unimodal Gaussian model cannot be accepted as an appropriate representation of the distribution
(see the AICc and $P$-values in Table~\ref{VelPVFdisTest}).
The bimodal Gaussian distribution consists of two distinct groups.
The NV group is characterized by a lower PVF velocity ($\langle V^{\rm PVF}_{\rm Si} \rangle_{\rm NV} = $ 10600 km/s)
with a narrower dispersion (600 km/s) and contributes 68$\pm$4 per cent of the total distribution.
In contrast, the HV group exhibits a higher PVF velocity ($\langle V^{\rm PVF}_{\rm Si} \rangle_{\rm HV} =$ 12100 km/s)
with a broader dispersion (1300 km/s) and accounts for 32$\pm$4 per cent of the distribution.
The superposition of these two groups is illustrated by the black solid
line in the upper-left panel of Fig.~\ref{SNVeldis}.
It is important to emphasize that this PVF velocity distribution is derived from measurements
taken at phases that are relatively uniformly distributed across all three SN~Ia subclasses
within the ten-day interval (see the upper-right inset in Fig.~\ref{SNVeldis}).
The median phase for each SN~Ia subclass is $\approx 0.5$ day.

For normal SNe~Ia in Table~\ref{VelPVFdisTest},
our results are consistent with those of \citet{2013Sci...340..170W},
who reported an NV component with
$\langle V_{\rm Si} \rangle_{\rm NV} = 10800$ km/s
(600 km/s dispersion and $\sim62$ per cent weight)
and an HV component with
$\langle V_{\rm Si} \rangle_{\rm HV} = 13000$ km/s
(1400 km/s dispersion and $\sim38$ per cent weight).
For all SNe~Ia, our results are consistent with those of \citet{2020MNRAS.499.5325Z},
who found an NV component with
$\langle V_{\rm Si} \rangle_{\rm NV} = 11000$ km/s
(700 km/s dispersion and 64.6 per cent weight)
and an HV component with
$\langle V_{\rm Si} \rangle_{\rm HV} = 12300$ km/s
(1800 km/s dispersion and 35.4 per cent weight).
Finally, using a combined untargeted sample restricted to redshifts $z<0.09$
and a phase interval of $-5 \lesssim t \lesssim +5$ days,
\citet{2024MNRAS.532.1887P} reported
$\langle V_{\rm Si} \rangle_{\rm NV} = 10700\pm700$ km/s
and $\langle V_{\rm Si} \rangle_{\rm HV} = 11900\pm1700$ km/s,
with corresponding population fractions of 76 and 24 per cent.
While the weight of the NV group in our sample (68$\pm$4 per cent) appears smaller than that
reported by \citet{2024MNRAS.532.1887P}, a simple binary probability (binomial distribution) consideration
suggests that their one $\sigma$ uncertainty on the weight estimate should be comparable to ours,
since their sample size is of the same order ($\sim150$ SNe~Ia).
Consequently, the differences between the NV and HV population fractions in the two studies are
likely within one to two $\sigma$.
In addition, the population differences between the targeted nature of our sample and
the untargeted nature of that used by \citet{2024MNRAS.532.1887P} may also contribute to the observed,
not significant, discrepancies in the reported weights.

For the PVF velocity distribution of the \mbox{Ca\,{\sc ii}} line,
the $P$-values from the KS and AD tests (Table~\ref{VelPVFdisTest}) indicate that the null hypothesis
is not rejected for either fit.
This suggests that both the bimodal and unimodal Gaussian models provide statistically acceptable descriptions
of the observed velocity data.
However, the transition from the unimodal to the bimodal model results in an AICc increase of
5.1 for all SNe~Ia and 4.5 for the subsample of normal SNe~Ia.
These values imply that the additional parameters in the bimodal Gaussian model do not significantly improve
the fit and may lead to overfitting.
In other words, there is no strong evidence that the bimodal model provides a superior fit to
the \mbox{Ca\,{\sc ii}} PVF velocity distribution compared to the unimodal model
(see AICc values in Table~\ref{VelPVFdisTest}).

To quantify the effect of measurement uncertainties on the choice between unimodal and bimodal velocity distributions,
we perform a MC analysis in which each measured velocity is perturbed according to a Gaussian distribution
defined by its individual uncertainty.
For each of $10^4$ realisations, we recalculate the MLE unimodal and bimodal fits and their corresponding AICc values.
The resulting distribution of $\Delta$AICc is summarized by its median and 16th--84th percentile range,
which provide robust indicators of the preferred model.
We find that for \mbox{Si\,{\sc ii}} PVF, the bimodal model remains strongly favored with
Med$(\Delta {\rm AICc}) = 23.1^{+5.0}_{-4.6}$,
confirming that the evidence for two velocity components is insensitive to measurement errors.
For \mbox{Ca\,{\sc ii}} PVF, the Med$(\Delta {\rm AICc}) = 2.5^{+2.1}_{-1.7}$.
This result reinforces that \mbox{Si\,{\sc ii}} PVF velocities are intrinsically bimodal,
whereas \mbox{Ca\,{\sc ii}} PVF velocities are consistent with a unimodal distribution
(see Med($\Delta$AICc) values in Table~\ref{VelPVFdisTest}).

To assess the robustness of our results against the specific choice of measurement epoch within the adopted
ten-day phase interval, we repeat the above-described analysis by selecting,
for each SN~Ia, the measurement closest to $t = -3, -2, -1, +1, +2$ and $+3$ days,
whenever such spectra were available.
As noted in Section~\ref{samplered}, some SNe~Ia in our sample have multiple spectra within
the ten-day interval, allowing this test to be performed without altering the number of objects in the sample.
This analysis shows that the results remain stable and are practically identical to those
obtained for the data centered at $t = 0$.
Therefore, the precise choice of measurement epoch within the $\pm5$ day interval
does not significantly affect the outcomes of the fits.

\begin{figure}
\begin{center}$
\begin{array}{@{\hspace{0mm}}c@{\hspace{0mm}}}
\includegraphics[width=\hsize]{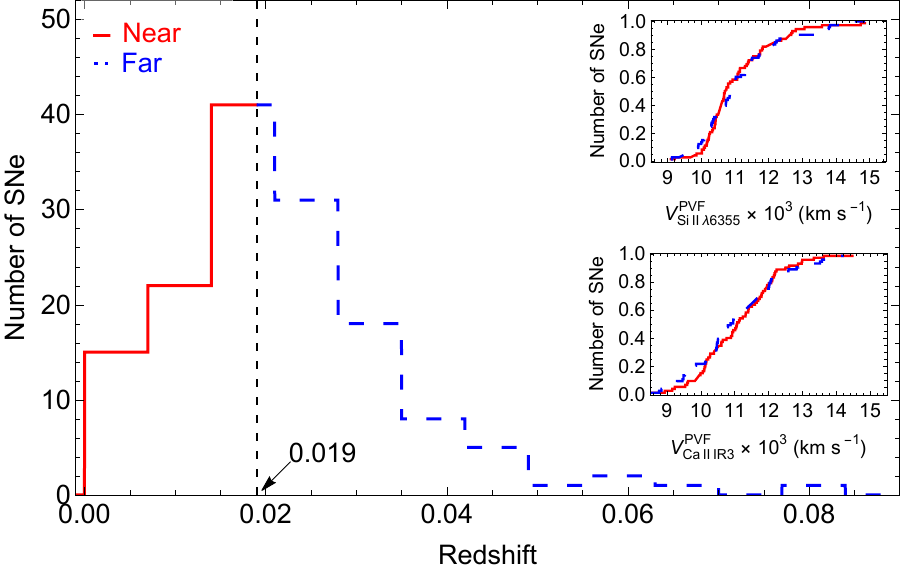}
\end{array}$
\end{center}
\caption{Histogram of redshifts for all SNe~Ia in our sample,
         divided into Near (red solid) and Far (blue dashed) subsamples
         using the median redshift (indicated by the vertical dashed line) as the division point.
         The median redshift of 0.019 is marked with an arrow.
         The upper and bottom insets display the CDFs of
         PVF velocities for the \mbox{Si\,{\sc ii}} and \mbox{Ca\,{\sc ii}} lines, respectively,
         separated according to the Near and Far subsamples.}
\label{SNredshiftHist}
\end{figure}
\begin{table}
  \caption{Comparison of the PVF velocity distributions of \mbox{Si\,{\sc ii}} and \mbox{Ca\,{\sc ii}}
           lines for SNe~Ia in our sample,
           separated into Near and Far subsamples based on redshift.}
  \label{VelNearFarTest}
  \centering
  \tabcolsep 11.7pt
    \begin{tabular}{lccccc}
    \hline
    \multicolumn{1}{l}{SN} & \multicolumn{1}{c}{$N_{\rm Near}$} & \multicolumn{1}{l}{vs} & \multicolumn{1}{c}{$N_{\rm Far}$} & \multicolumn{1}{c}{$P_{\rm KS}^{\rm MC}$} & \multicolumn{1}{c}{$P_{\rm AD}^{\rm MC}$}\\
    \hline
     \multicolumn{6}{c}{Si PVF velocities} \\
    All & 72 & vs & 73 & 0.633 & 0.647 \\
    Normal & 56 & vs & 57 & 0.783 & 0.771 \\
     \multicolumn{6}{c}{Ca PVF velocities} \\
    All & 72 & vs & 73 & 0.566 & 0.668 \\
    Normal & 56 & vs & 57 & 0.709 & 0.538 \\
   \hline
  \end{tabular}
  \parbox{\hsize}{\emph{Notes.} For the two-sample KS and AD tests, MC simulations with $10^4$ iterations are used to provide
             $P_{\rm KS}^{\rm MC}$ and $P_{\rm AD}^{\rm MC}$ probabilities.}
\end{table}

Another potential source of bias may affect the PVF velocity distributions presented in Fig.~\ref{SNVeldis}.
Since SN discoveries rely heavily on photometric surveys with limited depth and sensitivity,
and that HV SNe~Ia may, on average, differ in luminosity at maximum light
\citep[e.g.][]{2013Sci...340..170W,2018ApJ...858..104Z}
or exhibit distinct LC decline rates
\citep[e.g.][]{2014MNRAS.437..338C}
compared to their NV counterparts, HV SNe~Ia may preferentially be detected at segregated distances.
This potential selection effect could systematically distort the observed velocity distributions,
potentially either enhancing or underrepresenting the prominence of HV events.
To investigate this potential bias, following \citet{2020MNRAS.499.5325Z},
we divide our SN sample into two redshift-based groups: one at lower redshift (Near)
and the other at relatively higher redshift (Far).
We then perform two-sample KS and AD tests to compare the PVF velocity distributions of
Si (and Ca) between these two groups.
Fig.~\ref{SNredshiftHist} displays the redshift histogram of all SNe~Ia in our sample,
with the division point set at the median redshift.
The upper and bottom insets in this figure show the cumulative distribution functions (CDFs) of
PVF velocities for the \mbox{Si\,{\sc ii}} and \mbox{Ca\,{\sc ii}} lines, respectively,
separated according to the Near and Far subsamples.
For all SNe~Ia in our sample (and subset of normal SNe), the $P$-values from the KS and AD tests
indicate that the PVF velocity distributions of the Near and Far groups are statistically consistent
for both the Si and Ca lines (see Table~\ref{VelNearFarTest}).
Furthermore, to assess the robustness of this result,
we vary the redshift division point from the median value of 0.019 to alternative thresholds
at 0.016 and 0.022 and repeat the KS and AD tests.
In all cases, the resulting $P$-values remain above 0.3,
confirming the consistency of the velocity distributions across redshift-based groups.
Therefore, it is most likely that our sample does not exhibit significant redshift-dependent
biases in the PVF velocity distributions.
We note that recent studies indicate that any redshift evolution in the relative fractions of
HV and NV SNe~Ia is expected to become apparent only at $z > 0.05$ \citep[e.g.][]{2020ApJ...895L...5P,2024MNRAS.532.1887P},
which lies at the upper boundary of, and largely beyond, the redshift range covered by our sample.

\begin{figure}
\begin{center}$
\begin{array}{@{\hspace{0mm}}c@{\hspace{0mm}}}
\includegraphics[width=0.95\hsize]{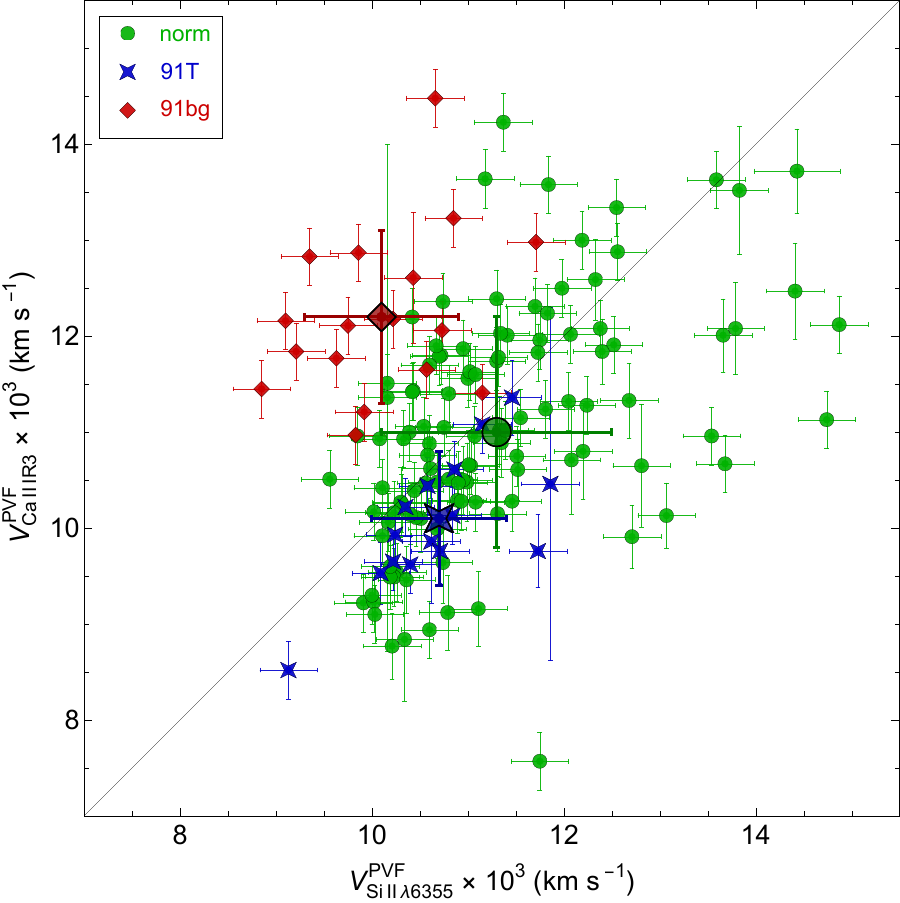}
\end{array}$
\end{center}
\caption{Comparison of PVF velocities of the \mbox{Si\,{\sc ii}}~$\lambda$6355 and
         \mbox{Ca\,{\sc ii}}~IR3 lines for SN~Ia subclasses.
         Mean $V^{\rm PVF}$ values for each subclass are indicated by larger symbols,
         with error bars representing their standard deviations.
         For reference, the diagonal line denotes the 1 to 1 relation.}
\label{VSivsVCa}
\end{figure}

To understand why the observed PVF velocity distributions of Si and Ca lines
behave notably differently, in the right panels of Fig.~\ref{SNVeldis}
we show the distributions of $V^{\rm PVF}_{\rm Si}$ and $V^{\rm PVF}_{\rm Ca}$,
separated by normal, 91T-, and 91bg-like SN~Ia subclasses.
As shown in the panels, the $V^{\rm PVF}_{\rm Si}$ distributions of 91T- and 91bg-like SNe
in our sample significantly overlap each other, being confined to
the lower end of the $V^{\rm PVF}_{\rm Si}$ distribution observed in normal SNe~Ia.
In contrast, the $V^{\rm PVF}_{\rm Ca}$ distributions for
91T- and 91bg-like SNe are clearly separated,
with the bulk of the $V^{\rm PVF}_{\rm Ca}$ values of normal SNe~Ia falling in between.
To further illustrate this behavior,
Fig.~\ref{VSivsVCa} presents a comparison of the PVF velocities of
the \mbox{Ca\,{\sc ii}} line versus those of the \mbox{Si\,{\sc ii}} line.
In Table~\ref{VelPVFetcSNtype}, we list the mean PVF velocities for both Si and Ca lines across different SN~Ia subclasses.
For normal and 91T-like SNe, the mean velocities of the two lines are statistically consistent
\citep[see also][]{2014MNRAS.444.3258M}.
However, a notable difference emerges for 91bg-like SNe: the \mbox{Ca\,{\sc ii}} line velocities exceed those of
the \mbox{Si\,{\sc ii}} line by $\sim2000~{\rm km~s}^{-1}$.
To assess the statistical significance of this discrepancy,
we conduct two-sample KS and AD tests comparing the $V^{\rm PVF}_{\rm Si}$ and $V^{\rm PVF}_{\rm Ca}$
distributions for the 91bg-like events,
with each measured velocity perturbed according to a Gaussian distribution defined by its individual uncertainty (as described above).
The resulting Med$(P_{\rm KS})$ and Med$(P_{\rm AD})$ values, both $<0.001$,
confirm that the distributions are significantly different ($>3.3\sigma$).

\begin{table}
  \caption{Mean PVF velocities $(\times 10^3)$ of \mbox{Si\,{\sc ii}} and \mbox{Ca\,{\sc ii}}
           lines and their standard deviations for different SN~Ia subclasses.}
  \label{VelPVFetcSNtype}
  \centering
  \tabcolsep 18.3pt
    \begin{tabular}{lccc}
    \hline
  \multicolumn{1}{l}{SN} & \multicolumn{1}{c}{$N_{\rm SN}$} & \multicolumn{1}{c}{$\langle V^{\rm PVF}_{\rm Si} \rangle$} & \multicolumn{1}{c}{$\langle V^{\rm PVF}_{\rm Ca} \rangle$} \\
  \hline
   Normal & 113 & 11.3$\pm$1.2 & 11.2$\pm$1.2 \\
   91T & 15 & 10.7$\pm$0.7 & 10.1$\pm$0.7 \\
   91bg & 17 & 10.1$\pm$0.8 & 12.2$\pm$0.9 \\
  \hline
  \end{tabular}
\end{table}

For normal SNe~Ia and 91T-like events, the results discussed above suggest that,
near maximum brightness, the PVF-forming layers for the \mbox{Si\,{\sc ii}}~$\lambda$6355
and \mbox{Ca\,{\sc ii}}~IR3 absorption lines are largely coincident.
In contrast, for 91bg-like SNe, the \mbox{Ca\,{\sc ii}} PVFs appear to form over
broader regions that, on average, trace the ejecta at significantly higher radii than
those associated with \mbox{Si\,{\sc ii}}.
This discrepancy is most likely attributed to differences in the photospheric (ejecta)
temperatures among SN~Ia subclasses \citep[e.g.][]{2005ApJ...623.1011B,2006MNRAS.370..299H},
as well as to the differing excitation energies required to produce the lines:
1.7~eV for \mbox{Ca\,{\sc ii}}~IR3 compared to 8.12~eV for \mbox{Si\,{\sc ii}}~$\lambda$6355.
In the relatively low temperature of the expanding photosphere of 91bg-like SNe,
the \mbox{Ca\,{\sc ii}}~IR3 absorption forms more easily and becomes saturated in the outer layers,
where the temperature is lower, whereas the \mbox{Si\,{\sc ii}}~$\lambda$6355 line
forms more efficiently in deeper, hotter layers \citep[e.g.][]{2015ApJS..220...20Z}.
As the photospheric temperature increases, such as in normal and 91T-like SNe,
the separation between the Si and Ca PVF-forming layers is expected to diminish.

\begin{figure}
\begin{center}$
\begin{array}{@{\hspace{0mm}}c@{\hspace{0mm}}}
\includegraphics[width=\hsize]{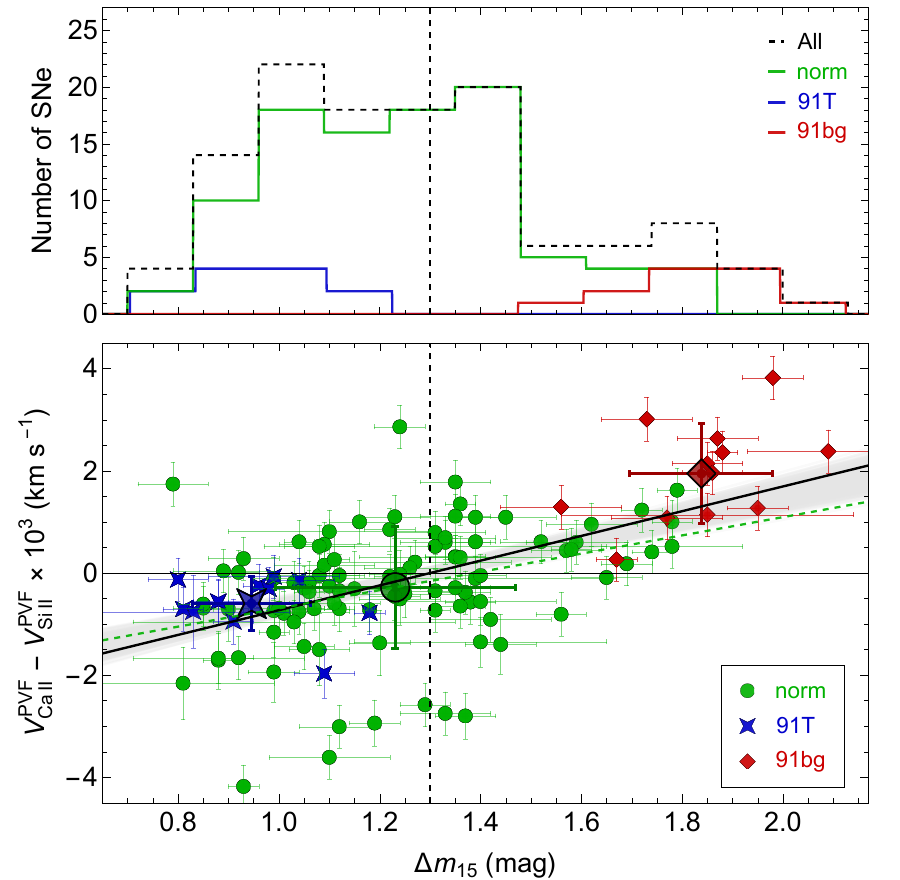}
\end{array}$
\end{center}
\caption{\emph{Upper panel:} Histograms of the LC decline rate $(\Delta m_{15})$ for SNe~Ia,
         separated by subclasses: normal (green), 91T-like (blue), and 91bg-like (red) events.
         The vertical dashed line at $\Delta m_{15}=1.3$ mag delineates the rough boundary for faster-declining SNe
         (see text for explanation).
         \emph{Bottom panel:} The difference between \mbox{Ca\,{\sc ii}} and \mbox{Si\,{\sc ii}} PVF velocities
         $(V^{\rm PVF}_{\rm Ca} - V^{\rm PVF}_{\rm Si})$ for the same SNe plotted against $\Delta m_{15}$.
         Mean values for each subclass are indicated by larger symbols,
         with error bars reflecting the standard deviations.
         Best linear fits for all SNe~Ia (solid black) and for normal events only (dashed green) are overlaid.
         For reference, the horizontal line denotes zero velocity difference.
         For all SNe, grey lines represent all $10^4$ linear fits to the MC realisations,
         illustrating the effect of measurement uncertainties on the correlation.}
\label{DVvsDm15fig}
\end{figure}
\begin{table}
  \caption{Results of Spearman's rank correlation tests for
           the $\Delta V^{\rm PVF}$ versus $\Delta m_{15}$ parameters
           of SNe~Ia.}
  \label{DVvsDm15Sptest}
  \centering
  \tabcolsep 8.2pt
    \begin{tabular}{lccccr}
    \hline
  \multicolumn{1}{l}{SN} & \multicolumn{1}{c}{$N_{\rm SN}$} & \multicolumn{1}{c}{$r_{\rm S}$} & \multicolumn{1}{c}{$P_{\rm S}^{\rm MC}$} & \multicolumn{1}{c}{Med$(r_{\rm S})$} & \multicolumn{1}{c}{Med$(P_{\rm S})$} \\
  \hline
   All & 121 & 0.54 & $<$\textbf{0.001} & $0.48^{+0.04}_{-0.04}$ & $<$\textbf{0.001} \\
   Normal & 97 & 0.39 & $<$\textbf{0.001} & $0.34^{+0.05}_{-0.05}$ & $\textbf{0.001}^{\textbf{+0.004}}_{\textbf{--0.001}}$ \\
  \hline
  \end{tabular}
  \parbox{\hsize}{\emph{Notes.} Spearman's coefficient $(r_{\rm S} \in [-1;1])$ is a non-parametric measure of
             rank correlation that assesses how well the relationship between two variables
             can be described using a monotonic function.
             Statistically significant correlations ($P \leq 0.05$) are highlighted in bold.}
\end{table}

Importantly, the LC decline rates of SNe~Ia serve as a useful proxy for photospheric temperature
\citep[e.g.][]{2006MNRAS.370..299H}.
Therefore, in the context of the above discussion,
analyzing the distribution of $\Delta V^{\rm PVF} \equiv V^{\rm PVF}_{\rm Ca} - V^{\rm PVF}_{\rm Si}$
versus $\Delta m_{15}$ for the different SN~Ia subclasses
can provide valuable insight (see Fig.~\ref{DVvsDm15fig}).
Table~\ref{DVvsDm15Sptest} presents the results of the Spearman's rank correlation test
between the $\Delta V^{\rm PVF}$ and $\Delta m_{15}$ variables.
The null hypothesis of the test states that the variables are independent,
while the alternative hypothesis posits that they are not.
For all and normal SN~Ia subsamples, the $P$-values of the test indicate that the null
hypothesis is rejected, suggesting a statistically significant monotonic relationship between
$\Delta V^{\rm PVF}$ and $\Delta m_{15}$.
Specifically, the test reveals a positive correlation,
and Fig.~\ref{DVvsDm15fig} indicates that larger differences between \mbox{Ca\,{\sc ii}} and \mbox{Si\,{\sc ii}}
PVF velocities are predominantly associated with faster-declining SNe~Ia,
which are generally expected to exhibit lower photospheric temperatures
\citep[e.g.][]{2005ApJ...623.1011B,2006MNRAS.370..299H}.
The upper panel of Fig.~\ref{DVvsDm15fig} presents the histograms of $\Delta m_{15}$ values
for the different SN~Ia subclasses, which are consistent with those reported for the
general SN~Ia population
\citep[e.g.][]{2013ApJ...770L...8P,2016MNRAS.460.3529A,2019MNRAS.486.5785S,2020MNRAS.499.1424H}.

To quantify the impact of measurement uncertainties on the correlations reported in Table~\ref{DVvsDm15Sptest},
both $\Delta V^{\rm PVF}$ and $\Delta m_{15}$ are perturbed according to Gaussian distributions defined
by their respective measurement errors.
We then generate $10^4$ random realisations of the dataset.
For each realisation, the Spearman coefficient and its corresponding $P$-value are computed.
The median and 16th--84th percentile ranges of these realisations are adopted as the final estimates
of the $r_{\rm S}$ and $P$-value (see Table~\ref{DVvsDm15Sptest}).
We find a median $r_{\rm S}$ of $0.48^{+0.04}_{-0.04}$ with a median $P_{\rm S} < 0.001$,
indicating a statistically significant positive correlation between
$\Delta V^{\rm PVF}$ and $\Delta m_{15}$ at a confidence level exceeding 99.9 per~cent
(corresponding to $\sim3.3\sigma$).
Among the $10^4$ MC realisations of the linear fits in Fig.~\ref{DVvsDm15fig},
none yield a negative slope, indicating that the sign of the correlation is fully robust.
The same holds true when only normal SNe Ia are considered.

\begin{figure}
\begin{center}$
\begin{array}{@{\hspace{0mm}}c@{\hspace{0mm}}}
\includegraphics[width=\hsize]{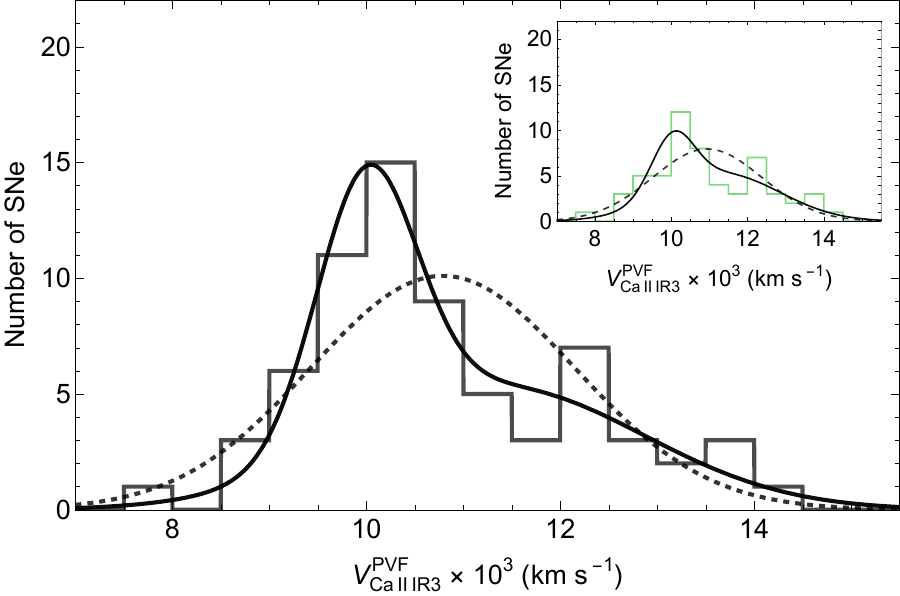}
\end{array}$
\end{center}
\caption{Distribution of \mbox{Ca\,{\sc ii}} PVF velocities for all SN~Ia subclasses (black histogram),
         restricted to events with $\Delta m_{15} \leq 1.3$ mag.
         The statistically preferred bimodal Gaussian fit is shown by the solid line,
         and the unimodal Gaussian fit by the dashed line.
         Inset shows the same distribution and fits restricted to normal SNe~Ia (green histogram).}
\label{disVCaLC13}
\end{figure}

To investigate whether the observed bimodal and unimodal behaviors in the PVF velocity distributions
of the \mbox{Si\,{\sc ii}} and \mbox{Ca\,{\sc ii}} lines, respectively,
depend on the LC decline rates, we repeat the velocity distribution analysis on
subsamples of SNe~Ia with $\Delta m_{15} \leq 1.2$ to $\Delta m_{15} \leq 1.7$ mag in steps of 0.05 mag.
Our results show that, across all these subsamples,
the PVF velocities of \mbox{Si\,{\sc ii}} consistently follow a bimodal Gaussian distribution,
similar to what is reported in Fig.~\ref{SNVeldis} and Table~\ref{VelPVFdisTest}.
In contrast, the PVF velocities of \mbox{Ca\,{\sc ii}} follow a bimodal Gaussian distribution
only for SNe~Ia with $\Delta m_{15} \leq 1.3$ mag (Med$(\Delta {\rm AICc}) > 6$);
beyond this threshold (delineated in Fig.~\ref{DVvsDm15fig}),
the distribution transitions to a unimodal Gaussian (Med$(\Delta {\rm AICc}) < 6$).
Fig.~\ref{disVCaLC13} illustrates the \mbox{Ca\,{\sc ii}} PVF velocity
distribution for SNe~Ia with $\Delta m_{15} \leq 1.3$ mag,
where the bimodal Gaussian model is statistically preferred over the unimodal one.

Thus, the most plausible explanation for the distinct behavior observed in the PVF velocity distributions
of the \mbox{Si\,{\sc ii}} and \mbox{Ca\,{\sc ii}} lines
(Fig.~\ref{SNVeldis} and Table~\ref{VelPVFdisTest}) is that faster-declining SNe~Ia tend to exhibit
higher \mbox{Ca\,{\sc ii}} PVF velocities, which,
on average, trace line-forming regions at higher ejecta layers
compared to those of \mbox{Si\,{\sc ii}}.
This effect arises due to the lower photospheric temperatures in such events,
particularly in normal SNe~Ia with $\Delta m_{15} > 1.3$ and 91bg-like SNe,
leading to stronger and more extended \mbox{Ca\,{\sc ii}} absorption.
In contrast, such an effect is not observed in SNe with $\Delta m_{15} < 1.3$,
which have relatively higher photospheric temperatures.
Consequently, when combining both slower and faster-declining SNe~Ia,
the cumulative effect shifts the \mbox{Ca\,{\sc ii}} PVF velocity distribution
from a bimodal toward a more unimodal Gaussian shape.

\subsection{Distributions of HVF velocities and strengths}
\label{RESsDIS2}

As mentioned in the Introduction, the high velocity absorption components
observed in the spectra of SNe~Ia are most likely produced by an absorbing shell
or circumstellar layer located exterior to the typical SN photosphere
(e.g. \citealt{2014MNRAS.437..338C}; \citetalias{2015MNRAS.451.1973S}).
Chronologically, the spectral evolution of SNe~Ia typically begins with the presence of
dominant \mbox{Ca\,{\sc ii}} HVFs.
Subsequently, PVFs strengthen alongside HVFs,
and finally, the HVFs diminish, leaving only PVFs
\citep[e.g.][]{2013ApJ...777...40M,2013ApJ...770...29C,2014MNRAS.437..338C}.
Importantly, the HVFs of \mbox{Si\,{\sc ii}}~$\lambda$6355 line disappear at earlier phases
than those of \mbox{Ca\,{\sc ii}}~IR3 line
(e.g. \citealt{2013ApJ...777...40M}; \citetalias{2015MNRAS.451.1973S}).
Therefore, HVFs associated with the \mbox{Si\,{\sc ii}} are relatively uncommon in SN~Ia
spectra (e.g. \citetalias{2015MNRAS.451.1973S}).
In our sample, restricted to a ten-day phase interval,
HVFs of the \mbox{Ca\,{\sc ii}} line are detected in $\sim 58$ per cent of normal SNe~Ia
(65 out of 113).\footnote{According to \citetalias{2015MNRAS.451.1973S},
if an HVF has a depth less than about 10 per cent of that of the PVF,
it cannot be reliably detected by their fitting algorithm.}
All 91T-like SNe, with the exception of one (14 out of 15),
exhibit HVFs, whereas only a single 91bg-like event (out of 17) shows such features.
\mbox{Si\,{\sc ii}} HVFs are present in only eight of the 145 SNe~Ia in our sample, all of which are spectroscopically normal events.
As for the PVFs, the HVF velocities decrease over time as the SN ejecta expand.
Along the time axis, a clear separation is observed between the \mbox{Ca\,{\sc ii}} HVF and PVF velocity components
(e.g. \citetalias{2015MNRAS.451.1973S}; \citealt{2015ApJS..220...20Z}).
For all SNe in our sample, the mean velocity offset between HVFs and PVFs of \mbox{Ca\,{\sc ii}} line
is $8200\pm1400$~${\rm km~s}^{-1}$.
These statistics align well with previously established results in the behavior of HVFs in SNe~Ia,
as discussed in \citet{2014MNRAS.437..338C,2014MNRAS.444.3258M,2015ApJS..220...20Z}.

To quantitatively characterize the presence of HVFs,
their strength relative to PVFs is typically assessed using the ratio of
pEWs of the respective components, defined as
$R={\rm pEW}_{\rm HVF}/{\rm pEW}_{\rm PVF}$.
This diagnostic ratio was introduced by \citet{2014MNRAS.437..338C} for
the \mbox{Ca\,{\sc ii}}~IR3 line $(R_{\rm Ca})$,
and subsequently adopted by \citetalias{2015MNRAS.451.1973S} for the \mbox{Si\,{\sc ii}}~$\lambda$6355
absorption feature $(R_{\rm Si})$.
According to the data compiled in \citetalias{2015MNRAS.451.1973S},
across all examined phases ($-16 \leq t \leq +5$ days), the vast majority of spectra show that
$R_{\rm Si} \lesssim R_{\rm Ca}$.
Moreover, a considerable number of spectra exhibit $R_{\rm Si}=0$,
indicating the absence of detectable \mbox{Si\,{\sc ii}} HVFs, whereas
$R_{\rm Ca}>1$ at the same epochs, signifying that the \mbox{Ca\,{\sc ii}} HVFs are stronger than their corresponding PVFs.
Therefore, in the remainder of this study, we adopt $R_{\rm HVF} \equiv R_{\rm Ca}$ as the primary parameter
to characterize HVF strength.

\begin{figure*}
\begin{center}$
\begin{array}{@{\hspace{0mm}}c@{\hspace{0mm}}}
\includegraphics[width=\hsize]{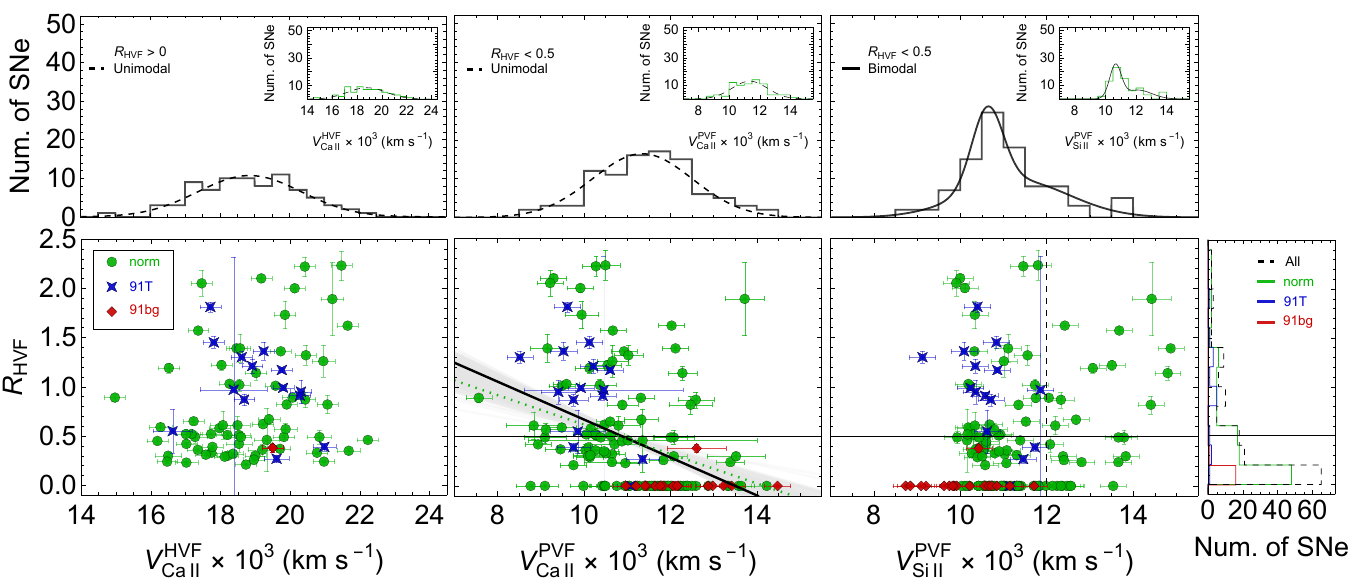}
\end{array}$
\end{center}
\caption{\emph{Upper panel:} Distributions of \mbox{Ca\,{\sc ii}} HVF and PVF velocities, as well as \mbox{Si\,{\sc ii}} PVF velocities
               for all SN~Ia subclasses (black histograms), restricted by their corresponding $R_{\rm HVF}$ values.
               Only the statistically preferred unimodal or bimodal Gaussian fits to the distributions are presented by dashed and solid thick lines, respectively.
               Insets show the same distributions and fits restricted to normal SNe~Ia (green histograms).
         \emph{Bottom panel:} Distribution of $R_{\rm HVF}$ as a function of \mbox{Ca\,{\sc ii}} HVF and PVF velocities,
               as well as \mbox{Si\,{\sc ii}} PVF velocities.
               The best linear fits for the statistically significant correlation between $R_{\rm HVF}$ and $V^{\rm PVF}_{\rm Ca}$
               are presented by solid black and dashed green thick lines for all and normal SN~Ia subclass, respectively.
               For all SNe, grey lines represent all $10^4$ linear fits to the MC realisations,
               illustrating the effect of measurement uncertainties on the correlation.
               The horizontal line marks $R_{\rm HVF}=0.5$, while the vertical dashed line separates NV and HV SNe~Ia.
               The histogram on the right shows the $R_{\rm HVF}$ distribution by SN~Ia subclasses.}
\label{histRCaALL}
\end{figure*}

For SNe~Ia in our sample, following \citet{2014MNRAS.444.3258M},
Fig.~\ref{histRCaALL} presents the distribution of $R_{\rm HVF}$ as a function of
the \mbox{Ca\,{\sc ii}} HVF and PVF velocities, as well as the \mbox{Si\,{\sc ii}} PVF velocities.
The figure also includes the corresponding histograms of $R_{\rm HVF}$ and $V^{\rm HVF}_{\rm Ca}$.
In addition, the histograms of $V^{\rm PVF}_{\rm Ca}$ and $V^{\rm PVF}_{\rm Si}$
are shown specifically for HVF-weak SNe~Ia, which we define as events with $R_{\rm HVF}<0.5$.
This threshold is adopted from \citet{2015ApJS..220...20Z},
although \citet{2014MNRAS.437..338C} defined HVF-weak SNe~Ia as those with $R_{\rm HVF}<0.2$.
It is important to note that this cut-off between HVF-weak and HVF-strong events is somewhat arbitrary.
Therefore, in subsequent analysis, we also explore the impact of varying this threshold within
the range of 0.2 to 0.5 values.

\begin{table}
  \caption{Results of Spearman's rank correlation tests for
           the $R_{\rm HVF}$ versus HVF or PVF velocities
           of SNe~Ia.}
  \label{RHVFvsHVFPVFvel}
  \centering
  \tabcolsep 2.3pt
    \begin{tabular}{lccccrrrr}
    \hline
  \multicolumn{1}{l}{SN} & \multicolumn{1}{c}{$N_{\rm SN}$} & \multicolumn{1}{c}{$R_{\rm HVF}$} & \multicolumn{1}{c}{vs} & \multicolumn{1}{c}{$V$} & \multicolumn{1}{c}{$r_{\rm S}$} & \multicolumn{1}{c}{$P_{\rm S}^{\rm MC}$} & \multicolumn{1}{c}{Med$(r_{\rm S})$} & \multicolumn{1}{c}{Med$(P_{\rm S})$}\\
  \hline
   All & 80 & \multicolumn{1}{c}{$R_{\rm HVF}$} & \multicolumn{1}{c}{vs} & \multicolumn{1}{c}{$V^{\rm HVF}_{\rm Ca}$} & 0.21 & 0.067 & $0.20^{+0.03}_{-0.03}$ & $0.074^{+0.058}_{-0.036}$ \\
   All & 145 & \multicolumn{1}{c}{$R_{\rm HVF}$} & \multicolumn{1}{c}{vs} & \multicolumn{1}{c}{$V^{\rm PVF}_{\rm Ca}$} & -- 0.53 & $<$\textbf{0.001} & $-0.51^{+0.02}_{-0.02}$ & $<$\textbf{0.001} \\
   All & 145 & \multicolumn{1}{c}{$R_{\rm HVF}$} & \multicolumn{1}{c}{vs} & \multicolumn{1}{c}{$V^{\rm PVF}_{\rm Si}$} & 0.07 & 0.391 & $0.08^{+0.03}_{-0.03}$ & $0.322^{+0.197}_{-0.141}$ \\
   Normal & 65 & \multicolumn{1}{c}{$R_{\rm HVF}$} & \multicolumn{1}{c}{vs} & \multicolumn{1}{c}{$V^{\rm HVF}_{\rm Ca}$} & 0.20 & 0.052 & $0.21^{+0.03}_{-0.03}$ & $0.058^{+0.030}_{-0.017}$ \\
   Normal & 113 & \multicolumn{1}{c}{$R_{\rm HVF}$} & \multicolumn{1}{c}{vs} & \multicolumn{1}{c}{$V^{\rm PVF}_{\rm Ca}$} & -- 0.43 & $<$\textbf{0.001} & $-0.42^{+0.03}_{-0.02}$ & $<$\textbf{0.001} \\
   Normal & 113 & \multicolumn{1}{c}{$R_{\rm HVF}$} & \multicolumn{1}{c}{vs} & \multicolumn{1}{c}{$V^{\rm PVF}_{\rm Si}$} & 0.05 & 0.575 & $0.07^{+0.03}_{-0.03}$ & $0.444^{+0.215}_{-0.166}$ \\
  \hline
  \end{tabular}
  \parbox{\hsize}{\emph{Notes.} The explanations for the $r_{\rm S}$ and $P$-values are the same as those provided in Table~\ref{DVvsDm15Sptest}.}
\end{table}

As shown in the bottom panels of Fig.~\ref{histRCaALL},
there is no apparent correlation between the $R_{\rm HVF}$ and either
the \mbox{Ca\,{\sc ii}} HVF or \mbox{Si\,{\sc ii}} PVF velocities.
SNe~Ia span a wide range of expansion velocities regardless of their $R_{\rm HVF}$ values.
In contrast, a negative trend may be apparent between $R_{\rm HVF}$ and \mbox{Ca\,{\sc ii}} PVF velocities,
suggesting that stronger HVFs could be associated with slower photospheric components.
This trend is supported by the results of the Spearman's rank correlation test presented in Table~\ref{RHVFvsHVFPVFvel}.
Among the tested combinations, a statistically significant correlation ($>3.3\sigma$) is found
only between $R_{\rm HVF}$ versus \mbox{Ca\,{\sc ii}} PVF velocities,
both for all SN~Ia sample and for the subsample of normal SNe~Ia.
Among the $10^4$ MC realisations of the linear fits in the bottom second panel of Fig.~\ref{histRCaALL},
none yield a positive slope, again demonstrating that the sign of the correlation is fully robust.
The same conclusion holds when only normal SNe~Ia are considered.

\begin{table}
  \caption{AICc and $\Delta$AICc values used for model selection between bimodal and unimodal Gaussian fits
           applied to the \mbox{Si\,{\sc ii}} and \mbox{Ca\,{\sc ii}} PVF velocity distributions of all and normal SNe~Ia,
           restricted by their corresponding $R_{\rm HVF}$ values.
           The results for the \mbox{Ca\,{\sc ii}} HVF velocity distribution are also presented.}
  \label{VelPVFHVFdisTest}
  \centering
  \tabcolsep 7.1pt
    \begin{tabular}{lcccrr}
    \hline
   & & \multicolumn{1}{c}{Bimodal} & \multicolumn{1}{c}{Unimodal} & \\
  \multicolumn{1}{l}{SN} & \multicolumn{1}{c}{$N$} & \multicolumn{1}{c}{AICc} & \multicolumn{1}{c}{AICc} & \multicolumn{1}{c}{$\Delta$AICc} & \multicolumn{1}{c}{Med($\Delta$AICc)} \\
   \hline
  \multicolumn{6}{c}{Si PVF velocities $(R_{\rm HVF}<0.5)$} \\
   All & 95 & 271.5 & 279.8 & \textbf{8.3} & $\textbf{6.3}^{\textbf{+4.6}}_{\textbf{--3.5}}$ \\
   Normal & 75 & 192.3 & 212.3 & \textbf{20.1} & $\textbf{14.1}^{\textbf{+5.1}}_{\textbf{--5.0}}$ \\
   \multicolumn{6}{c}{Ca PVF velocities $(R_{\rm HVF}<0.5)$} \\
   All & 95 & 304.5 & 302.1 & 2.4 & $2.5^{+2.2}_{-1.8}$ \\
   Normal & 75 & 238.8 & 238.7 & 0.1 & $2.5^{+2.4}_{-1.7}$ \\
   \multicolumn{6}{c}{Ca HVF velocities $(R_{\rm HVF}>0)$} \\
   All & 80 & 300.2 & 295.1 & 5.1 & $2.7^{+1.9}_{-1.8}$ \\
   Normal & 65 & 250.7 & 246.6 & 4.1 & $2.7^{+1.8}_{-1.7}$ \\
  \hline
  \end{tabular}
  \parbox{\hsize}{\emph{Notes.} The explanation of the AICc values is the same as in Table~\ref{VelPVFdisTest}.}
\end{table}

Possible correlations between $R_{\rm HVF}$ and either HVF or PVF velocities
have been previously investigated by several studies
(e.g. \citealt{2014MNRAS.444.3258M,2014MNRAS.437..338C}; \citetalias{2015MNRAS.451.1973S}; \citealt{2015ApJS..220...20Z}).
As shown in Table~\ref{RHVFvsHVFPVFvel},
our results confirm the absence of a statistically significant correlation between
$R_{\rm HVF}$ and $V^{\rm HVF}_{\rm Ca}$, consistent with the findings of \citet{2014MNRAS.444.3258M},
who reported no statistically significant difference in the mean $V^{\rm HVF}_{\rm Ca}$
values between their SNe~Ia with strong versus weak HVFs.
As already shown in \citetalias{2015MNRAS.451.1973S},
whose measurements constitute the basis of our dataset,
there is no evidence for a correlation between
$R_{\rm HVF}$ and $V^{\rm PVF}_{\rm Si}$,
nor any significant distinction in $R_{\rm HVF}$ between the HV and NV subgroups.
Accordingly, we show that the Spearman rank correlation tests between
$R_{\rm HVF}$ and \mbox{Si\,{\sc ii}} PVF velocities
all yield Med$(P_{\rm S}) > 0.3$, indicating no statistically significant relationship.
Moreover, the median $R_{\rm HVF}$ values for HV and NV SNe~Ia are
$0.42^{+0.80}_{-0.42}$ and $0.29^{+0.60}_{-0.29}$, respectively,
which are consistent within the 16th--84th percentile ranges.
This stands in contrast to the earlier conclusions of \citet{2014MNRAS.437..338C} and \citet{2014MNRAS.444.3258M},
who reported that HV SNe~Ia almost entirely lack detectable HVFs.
As noted by \citetalias{2015MNRAS.451.1973S},
many HV SNe~Ia do in fact exhibit moderate to strong HVFs near maximum light
(see quadrants in the bottom-third panel of Fig.~\ref{histRCaALL}).
The discrepancy with earlier findings is likely due to differences in
sample demographics, as previously discussed by \citetalias{2015MNRAS.451.1973S}.
Both \citet{2014MNRAS.437..338C} and \citet{2014MNRAS.444.3258M} had relatively few
HV events in their samples ($\sim13$ per cent), whereas our sample includes $\sim25$ per cent HV SNe~Ia,
comparable to the $\sim28$ per cent reported in \citetalias{2015MNRAS.451.1973S}.

On the other hand, our finding of a statistically significant negative correlation between
$R_{\rm HVF}$ and \mbox{Ca\,{\sc ii}} PVF velocities
(Table~\ref{RHVFvsHVFPVFvel} and the second panel from the bottom in Fig.~\ref{histRCaALL})
is consistent with the results of \citet{2014MNRAS.444.3258M},
who reported that SNe~Ia with stronger HVFs (i.e. larger $R_{\rm HVF}$ values) tend to exhibit
lower velocities in the \mbox{Ca\,{\sc ii}} near-IR photospheric component.
This raises the question: could the above discussed difference in the PVF velocity distributions,
bimodal for \mbox{Si\,{\sc ii}} and unimodal for \mbox{Ca\,{\sc ii}}
(Fig.~\ref{SNVeldis} and Table~\ref{VelPVFdisTest}),
be influenced by the inclusion of HVF-strong SNe~Ia in the sample,
for which the measured photospheric velocity components may
systematically vary with the $R_{\rm HVF}$ parameter?
To investigate this, we apply both bimodal and unimodal Gaussian models to the PVF velocity distributions
of \mbox{Si\,{\sc ii}} and \mbox{Ca\,{\sc ii}},
but restrict the analysis to the subset of HVF-weak SNe~Ia ($R_{\rm HVF}<0.5$; see Table~\ref{VelPVFHVFdisTest}).
Our results show that the conclusions presented in Section~\ref{RESsDIS1} remain unchanged when
HVF-strong events are excluded. The upper-second and third panels of Fig.~\ref{histRCaALL} display
the statistically acceptable unimodal and bimodal Gaussian fits for the Ca and Si PVF velocity distributions,
respectively, limited to HVF-weak SNe~Ia.
We therefore conclude that the strength of HVFs does not introduce a statistically significant bias
in the observed distributions of SN~Ia PVF velocities.

In addition, we examine the distribution of \mbox{Ca\,{\sc ii}} HVF velocities for SNe~Ia with $R_{\rm HVF}>0$
(see the upper-left panel in Fig.~\ref{histRCaALL}).
As shown in Table~\ref{VelPVFHVFdisTest}, the $V^{\rm HVF}_{\rm Ca}$ distribution
is well described by a unimodal Gaussian model.
The corresponding AICc values show no strong statistical preference for the bimodal model
for both all SN~Ia sample and the subsample of normal SNe~Ia with detectable HVFs.

To further validate our findings, we repeat the analysis presented in Table~\ref{RHVFvsHVFPVFvel}
using measurements obtained at phases closest to
$t= -3, -2, -1, +1, +2$, and $+3$ days, and compare these results to those centered at $t = 0$ days.
Similarly, we repeat the Gaussian fitting procedures described in Table~\ref{VelPVFHVFdisTest}
for both PVF and HVF velocity distributions using measurements at the same phase offsets.
In parallel, we vary the $R_{\rm HVF}$ threshold between 0.2 and 0.5 in steps of 0.1
to assess the impact of threshold selection.
In all configurations, the results remain stable and consistent with those derived from the baseline selections
(i.e., $t = 0$ for Table~\ref{RHVFvsHVFPVFvel}, and
$t = 0$, $R_{\rm HVF}<0.5$ for PVFs, and $R_{\rm HVF}>0$ for HVFs in Table~\ref{VelPVFHVFdisTest}).
These consistency checks demonstrate that our conclusions are not significantly influenced by either the adopted
$R_{\rm HVF}$ threshold or the specific timing of the velocity measurements within the ten-day phase interval.

\subsection{Dependence of Si PVF velocities on LC decline rates}
\label{RESsDIS3}

To investigate the potential connection between the SN~Ia ejecta kinematics and photometric evolution,
we examine the relationship between the \mbox{Si\,{\sc ii}}~$\lambda$6355 PVF velocity
near maximum light and the LC decline rate across different SN~Ia subclasses.
In particular, we assess this correlation while accounting for the level of HVF presence,
quantified by the $R_{\rm HVF}$ parameter.
Given that the kinetic energy of the explosion is expected to be directly related to
the expansion velocity of the SN ejecta
\citep[e.g.][]{2005ApJ...623.1011B,2006MNRAS.370..299H},
exploring this dependence may provide insights into how the photosphere interacts with the surrounding material.
Such interaction may involve CSM, swept-up material or shells
\citep[e.g.][]{2005MNRAS.357..200M,2008ApJ...677..448T,2019MNRAS.484.4785M},
potentially modifying the observable velocity field through possible energy exchange,
including density or abundance enhancement and ionization effects
(see \citealt{2019ApJ...886...58M} and references therein).
This analysis thus offers a path toward better understanding the diversity in SN~Ia explosion environments and
the role of HVFs in shaping observed photospheric velocities.

Notably, we exclude \mbox{Ca\,{\sc ii}} velocities
from analysis in this section, as they probe different ejecta layers depending on the SN~Ia subclass.
For instance, in 91bg-like and faster-declining normal SNe~Ia,
\mbox{Ca\,{\sc ii}} photospheric lines trace velocity regions
at significantly higher radii compared to those observed in 91T-like and slower-declining normal events (see Section~\ref{RESsDIS1}).
This strong subclass-dependent stratification of the Ca~II absorption lines complicates direct comparison of
$V^{\rm PVF}_{\rm Ca}$ or $V^{\rm HVF}_{\rm Ca}$ with $\Delta m_{15}$.

\begin{figure}
\begin{center}$
\begin{array}{@{\hspace{0mm}}c@{\hspace{0mm}}}
\includegraphics[width=0.95\hsize]{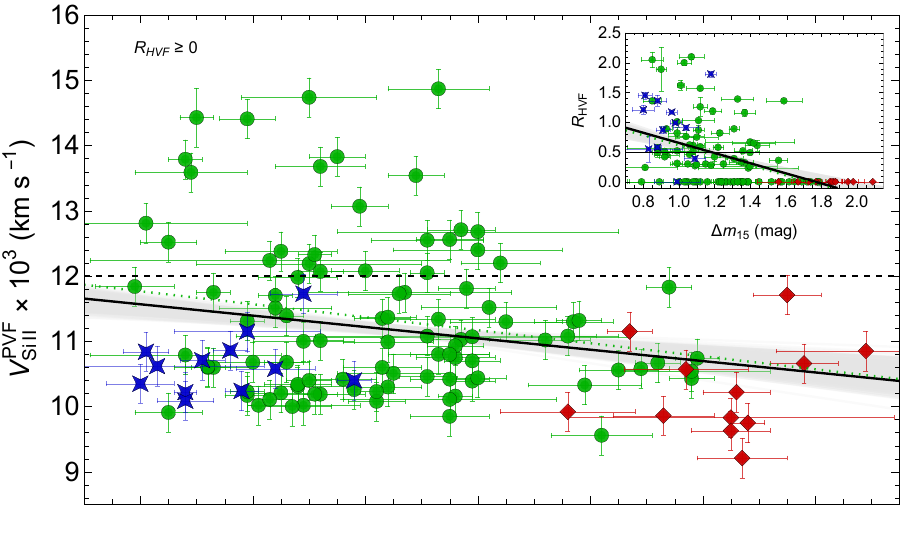} \\
\includegraphics[width=0.95\hsize]{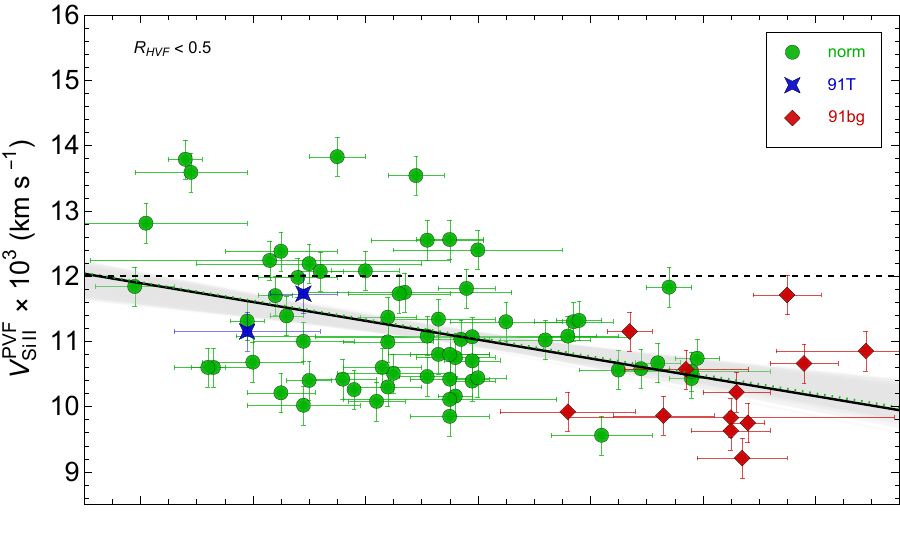} \\
\includegraphics[width=0.95\hsize]{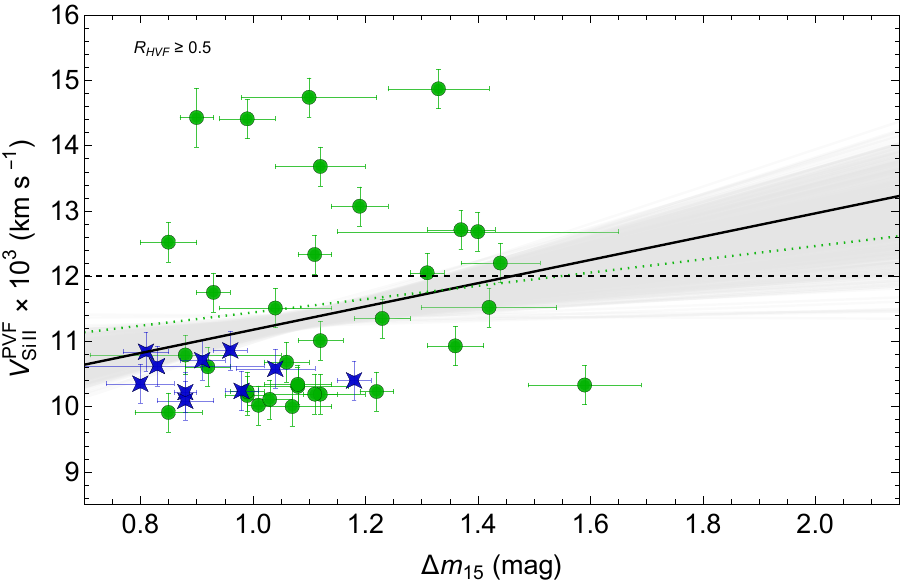} \\
\end{array}$
\end{center}
\caption{\emph{Upper panel:} \mbox{Si\,{\sc ii}}~$\lambda$6355 PVF velocity at $B$-band maximum light
         versus LC decline rate for different SN~Ia subclasses.
         The horizontal dashed line indicates the division between HV and NV events.
         The inset displays the relationship between $R_{\rm HVF}$ and $\Delta m_{15}$.
         The best linear fits between the mentioned parameters are presented by solid black and dashed green
         lines for all and normal SN~Ia subclass, respectively.
         For all SNe, grey lines represent all $10^4$ linear fits to the MC realisations,
         illustrating the effect of measurement uncertainties on the correlation.
         The horizontal line at $R_{\rm HVF}=0.5$ marks the division between HVF-weak and HVF-strong events.
         \emph{Middle and Bottom panels:} Same as the upper panel but restricted to
         HVF-weak and HVF-strong SNe~Ia, respectively.}
\label{SNVSivsdm15}
\end{figure}
\begin{table}
  \caption{Results of Spearman's rank correlation tests for
           the \mbox{Si\,{\sc ii}} PVF velocities
           versus $\Delta m_{15}$ values of SNe~Ia.
           The correlations are evaluated separately for all $(R_{\rm HVF} \geq 0)$,
           HVF-weak $(R_{\rm HVF}<0.5)$, and HVF-strong $(R_{\rm HVF}\geq0.5)$ subsets.}
  \label{VSivsDm15andRHVF}
  \centering
  \tabcolsep 5.2pt
    \begin{tabular}{lccrrrr}
    \hline
  \multicolumn{1}{l}{SN} & \multicolumn{1}{l}{$R_{\rm HVF}$} & \multicolumn{1}{c}{$N_{\rm SN}$} & \multicolumn{1}{c}{$r_{\rm S}$} & \multicolumn{1}{c}{$P_{\rm S}^{\rm MC}$} & \multicolumn{1}{c}{Med$(r_{\rm S})$} & \multicolumn{1}{c}{Med$(P_{\rm S})$} \\
  \hline
   All & $\geq 0$ & 121 & -- 0.13 & 0.149 & $-0.14^{+0.04}_{-0.04}$ & $0.134^{+0.155}_{-0.081}$ \\
   Normal & $\geq 0$ & 97 & -- 0.08 & 0.423 & $-0.10^{+0.05}_{-0.05}$ & $0.315^{+0.269}_{-0.172}$ \\
   All & $<0.5$ & 78 & -- 0.41 & $<$\textbf{0.001} & $-0.37^{+0.05}_{-0.05}$ & $\textbf{0.001}^{\textbf{+0.003}}_{\textbf{--0.001}}$ \\
   Normal & $<0.5$ & 64 & -- 0.27 & \textbf{0.029} & $-0.25^{+0.06}_{-0.06}$ & $\textbf{0.044}^{\textbf{+0.087}}_{\textbf{--0.032}}$ \\
   All & $\geq0.5$ & 43 & 0.30 & \textbf{0.048} & $0.25^{+0.07}_{-0.07}$ & $0.099^{+0.150}_{-0.066}$ \\
   Normal & $\geq0.5$ & 33 & 0.26 & 0.151 & $0.18^{+0.08}_{-0.08}$ & $0.318^{+0.270}_{-0.176}$ \\
  \hline
  \end{tabular}
  \parbox{\hsize}{\emph{Notes.} The explanations for the $r_{\rm S}$ and $P$-values are the same as those provided in Table~\ref{DVvsDm15Sptest}.}
\end{table}

In the upper panel of Fig.~\ref{SNVSivsdm15}, we present the distribution
of the \mbox{Si\,{\sc ii}} PVF velocity near maximum light as a function
of the LC decline rate for the various SN~Ia subclasses in our sample.
As noted in earlier studies
\citep[e.g.][]{2006MNRAS.370..299H,2009ApJ...699L.139W,2011ApJ...742...89F,2014MNRAS.437..338C},
there appears to be a dearth of HV events among fast-declining SNe~Ia.
In contrast, slow-declining SNe~Ia include HV events,
but there also appear to be a significant population of NV events.
At the same time, the results of the Spearman's rank correlation test, summarized in Table~\ref{VSivsDm15andRHVF},
show no statistically significant correlation between
$V^{\rm PVF}_{\rm Si}$ and $\Delta m_{15}$, either for all SN~Ia sample or for the subsample of normal events.
This finding is consistent with previous studies,
which have also reported the lack of a significant trend in the relationship between
$V^{\rm PVF}_{\rm Si}$ and LC decline rate
\citep[e.g.][]{1994AJ....108.2233W,2006MNRAS.370..299H,2014MNRAS.444.3258M,2021ApJ...923..267D,2024MNRAS.532.1887P}.

As discussed in detail by \citet{2014MNRAS.437..338C},
the observed photospheric velocity of the \mbox{Si\,{\sc ii}} line in SN~Ia
ejecta can be interpreted within the framework of both sub-Chandrasekhar double-detonation models
and Chandrasekhar-mass delayed-detonation scenarios
\citep[see][for recent reviews of different models]{2023RAA....23h2001L,2025A&ARv..33....1R}.
These explosion models predict stratified ejecta in which radioactive
$^{56}$Ni is synthesized in the inner regions,
while IMEs, including silicon, are formed at higher radii.
As a result, the measured $V^{\rm PVF}_{\rm Si}$ effectively traces the interface between
the IME-rich and $^{56}$Ni-rich layers.
In this context, a more energetic explosion, typically associated with low $\Delta m_{15}$
(i.e slow-declining events), is expected to show a higher $V^{\rm PVF}_{\rm Si}$,
as the IME--$^{56}$Ni boundary is pushed outward to larger radii
\citep[see also][]{2005ApJ...623.1011B,2006MNRAS.370..299H}.
However, the observation that SNe~Ia with low $\Delta m_{15}$
do not universally exhibit higher \mbox{Si\,{\sc ii}} velocities
(see the upper panel of Fig.~\ref{SNVSivsdm15})
challenges the predictive robustness of this stratification paradigm,
suggesting that additional factors may modulate the velocity structure in the ejecta.

\citet{2006MNRAS.370..299H} proposed that HVFs,
whether associated with \mbox{Ca\,{\sc ii}} or \mbox{Si\,{\sc ii}},
could serve as a contributing factor in modulating the observed velocity structure.
In particular, unresolved HVFs in the \mbox{Si\,{\sc ii}} absorption profile may lead to an artificial
overestimation of the photospheric velocity \citep[see also][]{2025A&A...695A.264H}.
Moreover, interaction between the photosphere and the outer HVF-forming layers could facilitate
different effects mentioned above thereby altering the dynamics of the expanding ejecta and
affecting the inferred PVF velocity.

As highlighted in previous studies
\citep[e.g.][]{2014MNRAS.444.3258M,2014MNRAS.437..338C,2015ApJS..220...20Z},
HVF-strong events are notably scarce among fast-declining SNe~Ia
(see the upper inset in Fig.~\ref{SNVSivsdm15}).
In contrast, slow-declining SNe~Ia encompass both HVF-strong and HVF-weak events,
indicating a broader diversity in the presence of HVFs among these objects.
The results of the Spearman's rank correlation test show
a statistically significant negative correlation $(>3.3\sigma)$
between $R_{\rm HVF}$ and $\Delta m_{15}$, with
Med$(r_{\rm S})=-0.52^{+0.03}_{-0.03}$ (Med$(P_{\rm S}) < 0.001$)
for all SN~Ia sample and
Med$(r_{\rm S})=-0.38^{+0.04}_{-0.03}$ (Med$(P_{\rm S}) < 0.001$)
for the subsample of normal events.
Among the $10^4$ MC realisations of the linear fits,
both for all SNe~Ia and for normal events only, none yield a positive slope.
Motivated by this trend and following \citet{2014MNRAS.437..338C},
we further investigate the relationship between $V^{\rm PVF}_{\rm Si}$ and $\Delta m_{15}$
by dividing the sample into HVF-weak and HVF-strong subsets
(see the middle and bottom panels of Fig.~\ref{SNVSivsdm15}).

The results of the Spearman's rank correlation test (Table~\ref{VSivsDm15andRHVF})
reveal a negative correlation between $V^{\rm PVF}_{\rm Si}$ and $\Delta m_{15}$ for HVF-weak SNe~Ia,
significant at the $>3.3\sigma$ level for all SNe~Ia
and at the $\sim2\sigma$ level for the subset of normal events
(see the middle panel of Fig.~\ref{SNVSivsdm15}).
None of the $10^4$ MC realisations produce a slope of the opposite sign,
for both all SNe~Ia and the subsample of normal events.
This supports the scenario in which more energetic explosions
(i.e slow-declining events) exhibit higher photospheric velocities.
The fact that this correlation is evident only for HVF-weak SNe~Ia
suggests that strong HVFs may mask or distort the underlying relationship
between explosion energetics and ejecta velocity.
This result aligns with the findings of \citet{2014MNRAS.437..338C},
who reported a similar trend between $V^{\rm PVF}_{\rm Si}$ and LC decline rate
for all SNe~Ia with $R_{\rm HVF}<0.2$.

On the other hand, for HVF-strong SNe~Ia, a marginally significant positive correlation $(\sim1.7\sigma)$
is observed between $V^{\rm PVF}_{\rm Si}$ and $\Delta m_{15}$ when
the SN~Ia subclass separation is not considered
(see Table~\ref{VSivsDm15andRHVF} and the bottom panel of Fig.~\ref{SNVSivsdm15}).
Among $10^4$ MC realisations of the linear fits, $\sim 0.07$ per cent yield a negative slope.
However, this correlation becomes statistically insignificant when the analysis is limited to normal SNe~Ia alone.
In this case, $\sim 2$ per cent of the $10^4$ MC realisations yield a slope of
the opposite sign, indicating that the inferred trend is not statistically stable.
Recall that the HVF-strong subset includes a substantial fraction of 91T-like objects,
as well as predominantly slow-declining normal SNe~Ia.
In this context, it is well-known that SNe~Ia originating from younger stellar populations,
particularly 91T-like events, which represent the youngest subclass of SNe~Ia
\citep[e.g.][]{2020MNRAS.499.1424H,2023MNRAS.520L..21B},
are more likely to be embedded in a dense CSM, in contrast to SNe~Ia arising from older environments
\citep[e.g.][]{2013MNRAS.431L..43J,2015A&A...574A..61L,2015ApJ...805..150F}.
This environmental distinction may contribute to the marginal trend observed for the HVF-strong subset
and its disappearance when 91T-like events are excluded.
At the same time, among spectroscopically normal SNe~Ia, the structure and density of the surrounding CSM
can vary considerably, reflecting differences in progenitor mass-loss histories, wind properties,
or the presence of circumstellar shells
\citep[e.g.][]{2007Sci...317..924P,2012Sci...337..942D,2019ApJ...886...58M}.
We note that, given the lack of direct CSM diagnostics in our dataset,
this explanation should be regarded as a plausible interpretation rather than a firm conclusion.

As in the previous section, we perform a series of robustness checks to
validate the findings presented in Table~\ref{VSivsDm15andRHVF}.
Specifically, we repeat the correlation tests using velocity measurements obtained at phases
closest to $t= -3, -2, -1, +1, +2$, and $+3$ days, while simultaneously varying the $R_{\rm HVF}$ threshold
between 0.2 and 0.5 in steps of 0.1.
In all corresponding configurations, the observed correlations remain stable and consistent with those derived from
the baseline selections presented in Table~\ref{VSivsDm15andRHVF} and Fig.~\ref{SNVSivsdm15}.
Again, these consistency checks confirm that our results are not significantly affected by either the choice of
$R_{\rm HVF}$ threshold or the precise timing of the velocity measurements within the adopted ten-day phase interval.

\section{Conclusions}
\label{CONCs}

This study revisits several established relations in SN~Ia ejecta kinematics and expands upon them
through a phase-matched reanalysis of the \citetalias{2015MNRAS.451.1973S} velocity dataset.
Using a ten-day phase interval centred on $B$-band maximum light,
we examine a well-defined and representative sample of 145 nearby SNe~Ia.
Building on previous studies, we introduce several new aspects:
(\emph{i}) the first direct, phase-matched comparison of
the \mbox{Si\,{\sc ii}}~$\lambda$6355 and \mbox{Ca\,{\sc ii}}~IR3 PVF
velocity distributions measured at identical epochs;
(\emph{ii}) an assessment of how the presence and strength of \mbox{Ca\,{\sc ii}} HVFs
influence these velocity distributions;
and (\emph{iii}) a re-evaluation of the correlation between
\mbox{Si\,{\sc ii}} PVF velocities and $\Delta m_{15}$,
explicitly accounting for HVF strength.
Together, these elements allow us to refine existing trends and identify
new physically motivated behaviours within the SN~Ia population.

We confirm that the \mbox{Si\,{\sc ii}} PVF velocity distribution exhibits a statistically significant bimodal structure,
in agreement with earlier studies \citep[e.g.][]{2020MNRAS.499.5325Z,2024MNRAS.532.1887P}, and
consistent with the well-established classification of normal SNe~Ia into NV and HV groups
\citep[e.g.][]{2013Sci...340..170W}.
For the first time, we demonstrate that the \mbox{Ca\,{\sc ii}} PVF velocity distribution,
measured for the same SNe~Ia at the same phases $(t({\rm Si}) = t({\rm Ca}))$,
is better represented by a unimodal Gaussian
(Fig.~\ref{SNVeldis} and Table~\ref{VelPVFdisTest}).
However, we emphasise that the \mbox{Ca\,{\sc ii}} PVF velocity distribution is bimodal under specific conditions,
most notably for SNe~Ia with $\Delta m_{15} \leq 1.3$ mag (Fig.~\ref{disVCaLC13}),
before transitioning to a unimodal form once higher-$\Delta m_{15}$ (cooler) events are included.
This contrast likely reflects a subclass-dependent formation depth of the Ca absorption line
(Fig.~\ref{VSivsVCa} and Table~\ref{VelPVFetcSNtype}),
as supported by a positive correlation $(>3.3\sigma)$
between LC decline rate ($\Delta m_{15}$) and the velocity offset between
\mbox{Ca\,{\sc ii}} and \mbox{Si\,{\sc ii}} PVFs
(Fig.~\ref{DVvsDm15fig} and Table~\ref{DVvsDm15Sptest}).
These results point to differences in excitation conditions, with Ca lines forming at higher radii in cooler,
lower-temperature ejecta such as in 91bg-like SNe~Ia and faster-declining normal events.

We also investigated the properties and statistical behavior of \mbox{Ca\,{\sc ii}} HVFs,
using the $R_{\rm HVF}$ ratio as a diagnostic of line strength.
In agreement with earlier studies \citep[e.g.][]{2014MNRAS.444.3258M},
a significant negative correlation $(>3.3\sigma)$
between $R_{\rm HVF}$ and $V^{\rm PVF}_{\rm Ca}$ suggests that stronger
HVFs preferentially occur in events with slower photospheric components
(Fig.~\ref{histRCaALL} and Table~\ref{RHVFvsHVFPVFvel}).
The HVF velocity distribution is well described by a unimodal Gaussian.
Importantly, we show that the presence of HVFs does not significantly bias
the observed bimodal and unimodal behaviors in the PVF velocity distributions of
the \mbox{Si\,{\sc ii}} and \mbox{Ca\,{\sc ii}} lines, respectively,
as confirmed through subset analysis restricted to HVF-weak events
(Fig.~\ref{histRCaALL} and Table~\ref{VelPVFHVFdisTest}).

Turning to the relationship between kinematic and photometric properties,
we explored how \mbox{Si\,{\sc ii}} PVF velocities depend on the LC decline rate $\Delta m_{15}$.
No significant correlation is confirmed for all SN~Ia sample,
but a statistically significant negative correlation
$(>3.3\sigma)$ emerges among HVF-weak events
(the middle panel of Fig.~\ref{SNVSivsdm15} and Table~\ref{VSivsDm15andRHVF}).
This result supports the general concept in which more energetic events (i.e. slow-declining SNe~Ia)
produce faster-expanding ejecta, with $V^{\rm PVF}_{\rm Si}$ tracing the IME--$^{56}$Ni boundary
\citep[see e.g.][]{2014MNRAS.437..338C}.
In contrast, for HVF-strong events, dominated by 91T-like and slow-declining normal SNe~Ia,
the correlation is absent or even inverted $(\sim1.7\sigma)$,
likely due to enhanced circumstellar interaction affecting the apparent PVF line velocities.
These findings are robust against measurement uncertainties,
variations in the specific timing of the velocity measurements
within the ten-day phase interval,
and differences in the adopted HVF strength thresholds.

Overall, our results underscore the importance of accounting for both spectroscopic subclass and
HVF strength when interpreting the kinematic and spectroscopic properties of SNe~Ia,
both in explosion models and observational data.
We encourage future studies to extend this work by incorporating broader temporal coverage,
additional spectral lines, direct CSM diagnostics, and a more diverse range of SN~Ia subclasses,
preferably discovered by untargeted surveys.
Such efforts will be essential for refining our understanding of SN~Ia explosion physics.

\section*{Acknowledgements}

We thank the anonymous referee for thoughtful comments and constructive suggestions,
which helped to improve the clarity and quality of this paper.
The research was supported by the Higher Education and Science
Committee of MESCS RA (Research project \textnumero~24LCG--1C021).
A.A.H. and A.G.K. acknowledge the hospitality of the
Institut d'Astrophysique de Paris (France) during their
stay as visiting scientists supported by
the Programme Visiteurs Ext\'{e}rieurs (PVE).
V.A. acknowledges support from FCT -- Funda\c{c}\~ao para a Ci\^encia e Tecnologia through
national funds and from FEDER via COMPETE2020 -- Programa Operacional Competitividade e Internacionaliza\c{c}\~ao,
under the grants UIDB/04434/2020 (DOI: 10.54499/UIDB/04434/2020) and UIDP/04434/2020 (DOI: 10.54499/UIDP/04434/2020),
as well as through a work contract funded by the FCT Scientific Employment Stimulus program
(reference 2023.06055.CEECIND/CP2839/CT0005, DOI: 10.54499/2023.06055.CEECIND/CP2839/CT0005).

\section*{Data Availability}

The names of all SNe~Ia in our sample, including their spectroscopic subclassifications and
$\Delta m_{15}$ values (with corresponding data sources), as well as their redshifts
are available in the online version (Supporting Information) of the article.
The PVF velocities and pEWs of the \mbox{Si\,{\sc ii}}~$\lambda$6355
and \mbox{Ca\,{\sc ii}}~IR3 absorption features used in this study, including the HVF properties of the \mbox{Ca\,{\sc ii}} line
and the phases at which the measurements were obtained,
are accessible in the online database of \citetalias{2015MNRAS.451.1973S}.



\bibliographystyle{mnras}
\bibliography{references} 



\section*{Supporting information}

Supplementary data are available at \emph{MNRAS} online.\\
\\
\textbf{supplementary.csv}\\
\\
Please note: Oxford University Press is not responsible for the
content or functionality of any supporting materials supplied by
the authors. Any queries (other than missing material) should be
directed to the corresponding author for the article.

%
%
%
%
%

\bsp	
\label{lastpage}
\end{document}